\documentclass[twocolumn,prd,floatfix,preprintnumbers,a4paper,nofootinbib,superscriptaddress,groupedaddress]{revtex4-1}
\usepackage[utf8]{inputenc}
\usepackage{amsmath,amssymb}
\usepackage{amsfonts}
\usepackage{bm}
\usepackage{courier}
\usepackage{graphics}
\usepackage{float}
\usepackage{amsmath}
\usepackage{amssymb}
\usepackage[normalem]{ulem} 
\usepackage[mathscr]{euscript}
\usepackage{acronym}
\usepackage{float}
\usepackage[caption=false]{subfig}
\usepackage{lipsum}
\usepackage{mathtools}
\usepackage{multirow}
\usepackage{graphicx}
\usepackage{multirow}
\usepackage{graphicx}
\usepackage[usenames,dvipsnames,table,xcdraw]{xcolor}
\usepackage[colorlinks, pdfborder={0 0 0}]{hyperref} 
\usepackage[nameinlink,capitalise]{cleveref}
\usepackage{tensor}
\usepackage{verbatim}
\usepackage{wrapfig}
\usepackage{soul} 

\graphicspath{{fig}}

\newcommand{\milan}{\affiliation{Dipartimento di Fisica ``G. Occhialini'', 
Universit\'a degli Studi di Milano-Bicocca, Piazza della Scienza 3, 20126 Milano, Italy}}
\newcommand{\infn}{\affiliation{INFN, Sezione di Milano-Bicocca, 
Piazza della Scienza 3, 20126 Milano, Italy}}

\date{\today}

\begin{document}
\title{Identifying modified theories of gravity using binary black-hole ringdowns}

\author{Costantino Pacilio}
\email{costantino.pacilio@unimib.it}
\milan
\infn

\author{Swetha Bhagwat}
\email{sbhagwat@star.sr.bham.ac.uk}
\affiliation{Institute for Gravitational Wave Astronomy $\&$ School of Physics and Astronomy, University of Birmingham,
Edgbaston, Birmingham B15 2TT, UK}

\begin{abstract}
Black-hole spectroscopy, that is, measuring the characteristic frequencies and damping times of different modes in a black-hole ringdown, is a powerful probe for testing deviations from the general theory of relativity (GR). In this work, we present a comprehensive study on its ability to identify deviations from the spectrum of a Kerr black hole in GR. Specifically, we investigate the performance of black hole spectroscopy on a diverse set of theoretically motivated as well as phenomenologically modified spectra. We find that while the signal-to-noise ratio $\rho_{\rm RD}$ in the ringdown required to identify a modification to the GR Kerr black hole spectrum depends on the details of the modifications, a modification that introduces $\sim 1 \%$ shift in the fundamental mode frequencies can typically be distinguished with $\rho_{\rm RD} \in [150,500]$. This range of $\rho_{\rm RD}$ is feasible with the next-generation detectors, showing a promising science case for black hole spectroscopy.
\end{abstract}

\maketitle
\section{Introduction}
\label{sec:intro}
Gravitational waves (GWs) with characteristic frequencies and damping times are radiated as the distorted black hole (BH) formed during a binary BH merger relaxes into its final stable state. This signal is called the ringdown and comprises of a linear superposition of the spectral modes of the BH, known as the quasi-normal modes (QNMs). We can obtain the QNMs by solving the BH perturbation equations \cite{Teukolsky:1972,Teukolsky:1973,Teukolsky:1974}  and the ringdown signal can be used to validate dynamics in linear strong field regime. The QNM spectra of a perturbed BH in the general theory of relativity (GR) are obtained by solving Teukolsky’s equation \cite{Pani:2013pma} and under the Kerr hypothesis \cite{BAMBI_2011,Teukolsky:2014vca} i.e., that the remnant BH in binary BH coalescence relaxes to a Kerr BH. 

BH spectroscopy \cite{Dreyer:2003bv}, defined as measuring QNM spectra from ringdown signals, allows to put forth consistency tests of the joint hypotheses that 
\begin{enumerate}
    \item the asymptotic equilibrium state of the remnant is described by the Kerr metric a.k.a., the Kerr hypothesis and
    \item the dynamics of the perturbed Kerr BH is governed by Teukolsky’s equation, i.e., (linearized) GR dynamics.
\end{enumerate}
Further, BH spectroscopy can observationally validate the no-hair theorem obeyed by BHs in GR; it demands that all aspects of a Kerr spacetime, including its QNM spectrum, be fully characterized by just two parameters. Often and most naturally, the two parameters are chosen as the mass $M_{f}$ and spin $\chi_{f}$ of the BH. If more than two QNM parameters are measured, it can be verified that different pairs of QNM parameters solve for the same $M_{f}$ and $\chi_{f}$. This  allows us to perform a null test of the no-hair theorem.

There has been much focus in the literature on the feasibility of measuring the subdominant QNM modes and performing null tests to validate the underlying theory of gravity as GR with BH spectroscopy \cite{Bhagwat:2021kwv,Bhagwat:inprep,Maggiore_2020,Berti:2016lat,Maselli:2017kvl,Carullo:2019flw,Bhagwat:2021kfa,Ghosh:2021mrv,Bhagwat:2019bwv,PhysRevD.105.044015,Baibhav:2018rfk,Baibhav:2020tma, Brito:2018rfr}. In this study, we concentrate on a complementary aspect and investigate the ability of BH spectroscopy to identify deviations from GR when the spectrum is not described by the GR Kerr QNMs. We perform a comprehensive study utilizing the publicly available QNM spectra in various modified theories as well as two phenomenological modifications and assess the performance of BH spectroscopy to differentiate them from a GR Kerr BH spectra. We then investigate the signal-to-noise ratio $\rho_{\rm RD}$ in the ringdown at which different modified theories of gravity can be distinguished from GR using BH spectroscopy. We find that the required $\rho_{\rm RD}$ depends on the details of the QNM spectra in a given theory and on their degeneracies with the GR Kerr BH spectra in the mass-spin space. However, at a broad level, we observe that $\rho_{\rm RD} \geq 150$ is required to confidently identify modified theories that produce $\leq 1 \%$ deviation in the dominant mode from GR using BH spectroscopy (c.f., \cite{Bhagwat:2021kwv,Bhagwat:inprep} for a detailed study on expected $\rho_{\rm RD}$ for measurability of QNM parameters with the next-generation GW detectors). 

The remainder of this paper is organized as follows. In Section \ref{sec:test} we outline the conceptual structure adopted for this study to test the no-hair theorem using BH spectroscopy. In Section \ref{sec:spectra}, we detail the modified QNM spectra used in this study. Then, in Section \ref{sec:methods} we summarize the setup and implementation used to perform this study. This is followed by the results in Section \ref{sec:results} and a discussion of their implications in Section \ref{sec:dis}.
\section{Testing the no-hair hypothesis with BH spectroscopy}
\label{sec:test}

The ringdown waveform observed at asymptotic infinity can be approximated as a linear superposition of a countably infinite set of (complex) QNMs with
\begin{equation}
    \omega_{lmn}=2\pi f_{lmn}-i/\tau_{lmn} .
\end{equation}
Here $f_{lmn}$ and $\tau_{lmn}$ are the characteristic frequencies and damping times of the spectral modes. $(l,m,n)$ index the mode's  angular, azimuthal and overtone numbers. As in any perturbation theory, the excitation amplitude of the modes depend on the initial perturbation conditions; for a binary BH merger these are set largely during the plunge-merger phase. A quasi-circular merger excites $(2,2,0)$ dominantly, and depending on the initial binary BH's mass ratio and spins, the most prominent subdominant angular modes can be $ \{ (3,3,0), (2,1,0), (4,4,0) \} $ \cite{Kamaretsos:2011um,Gossan:2011ha,Kamaretsos:2012bs,London:2014cma,Borhanian:2019kxt,MaganaZertuche:2021syq,Forteza:2020cve,Forteza:2022tgq,Borhanian:2019kxt}. 

To outline our setup, let us consider a ringdown where more than two QNM parameters are measurable, and a case where we have identified its QNM indices. The minimum $\rho_{\rm RD}$ required for this has been investigated in studies such as \cite{Berti:2007zu,Bhagwat:2019dtm,Bhagwat:2017tkm,London:2014cma}. If the underlying theory of gravity is GR and if the Kerr hypothesis holds, we can invert any pair of QNM parameters, preferably the frequency and damping time of the dominant mode, to infer the mass and spin of the BH --

\begin{equation}
\label{eq:invert}
\{f_{220},\tau_{220}\}\to\{{M_{f}^{\rm Kerr}},{\rm \chi_{f}^{\rm Kerr}}\}.
\end{equation}

From this mass and spin estimate, we can compute the full set of QNM spectra of the Kerr BH. Let $f_{lmn}^{\rm(infer)}$ be the inferred subdominant mode frequency 
\begin{equation}   
\label{eq:infer}   
\{{M_{f}^{\rm Kerr}},{\chi_{f}^{\rm Kerr}}\}\to f_{lmn}^{\rm(infer)}\,.
\end{equation}

Here we use the superscript (infer) to differentiate $f_{lmn}^{\rm(infer)}$ from  $f_{lmn}^{\rm(meas)}$ which are the frequencies measured from the ringdown signal.
While a similar argument holds for QNM damping times, subdominant mode damping times are poorly measured \cite{Gossan:2011ha, Berti:2005ys, Bhagwat:2021kwv} and therefore, we focus on tests using solely the subdominant mode frequencies.

A null test can be performed by checking if the relative difference between the inferred and the measured quantity is compatible with zero. We define the relative difference as

\begin{equation}   
\label{eq:Delta} 
\delta f_{lmn} = \frac{f_{lmn}^{\rm(meas)}-f_{lmn}^{\rm(infer)}}{f_{lmn}^{\rm(infer)}}\,.
\end{equation}

We can infer steps \eqref{eq:invert}-\eqref{eq:Delta} through a convenient reparametrization of the waveform during the parameter estimation. We briefly summarize this and point the reader to a detailed treatment in \cite{Isi:2021iql}.  

A generic modified QNM spectrum can be phenomenologically written as 
\begin{subequations}
\label{eq:mod:spectrum}
\begin{align}
    &f_{lmn}=f_{lmn}^{\rm Kerr}(M_f,\chi_f)(1+\delta f_{lmn})\,,\\
    &\tau_{lmn}=\tau_{lmn}^{\rm Kerr}(M_f,\chi_f)(1+\delta \tau_{lmn})\,,
\end{align}
\end{subequations}
where $\{M_f,\chi_f\}$ are the true values of the final mass and spin, and $\{\delta f_{lmn},\delta \tau_{lmn}\}$ are the relative shifts of the spectrum w.r.t.~the QNM spectra of a GR BH. $\{\delta f_{lmn},\delta \tau_{lmn}\}$ can be non-trivial functions of $\{M_f,\chi_f\}$ and of the physical parameters of the modified theory such as the additional coupling constants or charges. At this stage of setting up the formalism, we do not differentiate between a modification to the underlying theory of gravity and a modification in the nature of the compact object.  

Now, notice that $\{\delta f_{220},\delta \tau_{220}\}$ are redundant parameters because we can always find a pair $\{\tilde{M}_f,\tilde{\chi}_f\}$ of \textit{effective} final mass and spin that satisfy

\begin{subequations}
\label{eq:reparam:1}
\begin{align}
    &f_{220}=f_{220}^{\rm Kerr}(\tilde{M}_f,\tilde{\chi}_f)\,,
\label{eq:reparam:1:a}\\
    &\tau_{220}=\tau_{220}^{\rm Kerr}(\tilde{M}_f,\tilde{\chi}_f)\,.
\label{eq:reparam:1:b}
\end{align}
\end{subequations}
The subdominant modes can be re-expressed as

\begin{subequations}
\label{eq:reparam:2}
\begin{align}
    &f_{lmn}=f_{lmn}^{\rm Kerr}(\tilde{M}_f,\tilde{\chi}_f)(1+\tilde{\delta} f_{lmn})\,,
\label{eq:reparam:2:a}\\
    &\tau_{lmn}=\tau_{lmn}^{\rm Kerr}(\tilde{M}_f,\tilde{\chi}_f)(1+\tilde{\delta} \tau_{lmn})\,,
\label{eq:reparam:2:b}
\end{align}
\end{subequations}
for $(lmn)\neq(220)$. Further, the effective shifts $\{\tilde{\delta} f_{lmn},\tilde{\delta}\tau_{lmn}\}$ satisfy
\begin{subequations}
    \label{eq:reparam:3}
    \begin{align}
    & f_{lmn}^{\rm Kerr}(M_f,\chi_f)\left(1+\delta f_{lmn}\right) = f_{lmn}^{\rm Kerr}(\tilde{M}_f,\tilde{\chi}_f)(1+\tilde{\delta} f_{lmn})\,,
    \label{eq:reparam:3:a}\\
    & \tau_{lmn}^{\rm Kerr}(M_f,\chi_f)\left(1+\delta \tau_{lmn}\right) = \tau_{lmn}^{\rm Kerr}(\tilde{M}_f,\tilde{\chi}_f)(1+\tilde{\delta} \tau_{lmn})\,.
    \label{eq:reparam:3:b}
    \end{align}
\end{subequations}

For the QNM spectrum of a Kerr BH in GR, $\{\delta f_{lmn},\delta\tau_{lmn}\}$ vanish; therefore $\{\tilde{M}_f,\tilde{\chi_f}\}=\{M_f,\chi_f\}$ and $\{\tilde{\delta} f_{lmn},\tilde{\delta}\tau_{lmn}\}$ vanish. We set up our framework to identify departure from the GR Kerr BH QNM spectrum by constraining the effective shifts away from zero. 

Note that the mass and spin appearing in Eq.s~\eqref{eq:invert}-\eqref{eq:infer} are not the true values $\{M_f,\chi_f\}$ but rather the effective values $\{\tilde{M}_f,\tilde{\chi}_f\}$. We emphasise that we can only measure the \emph{effective} final mass and spin, and not the true values corresponding to the BHs. While developing a framework for observational test,  the effective (measured) parameters deviations $\{\tilde{\delta} f_{lmn},\tilde{\delta}\tau_{lmn}\}$ are the instrumental variables \footnote{Note also that if one is not interested in testing the no-hair theorem, the signals can be analyzed by assuming the Kerr BH spectrum (i.e., ~setting all $\{\delta f_{lmn},\delta\tau_{lmn}\}$ to zero) and recovering posterior estimates of the mass and spin. The posteriors so obtained will generally differ from the posteriors of $\{\tilde{M}_f,\tilde{\chi}_f\}$. This means that $\{\tilde{M}_f,\tilde{\chi}_f\}$ obtained here cannot be used to gauge the performance of tests like the Inspiral-merger-ringdown test \cite{Ghosh:2017gfp}.}. Similarly, the magnitudes of $\{\delta f_{lmn},\delta\tau_{lmn}\}$ are not directly accessible in BH spectroscopy and we can only estimate $\{\tilde{\delta} f_{lmn},\tilde{\delta}\tau_{lmn}\}$. In Section \ref{sec:spectra}, we inspect modified QNM spectra and show that $\{\tilde{\delta} f_{lmn},\tilde{\delta}\tau_{lmn}\}$ can be significantly different from $\{\delta f_{lmn},\delta\tau_{lmn}\}$.
\section{Modified QNM spectra}
\label{sec:spectra}

In this study, we quantify the ability of BH spectroscopy to constrain $\tilde{\delta}f_{lmn}$ away from zero for various class of modifications to the Kerr BH spectrum (c.f., \cite{Tattersall:2019pvx,Bao:2019kgt,Carullo:2021oxn,Silva:2022srr} for other works on BH spectroscopy in the context of modified theories of gravity). Given the lack of a best-candidate theory for modified gravity and the fact that QNM spectra in modified theories are available in a very few theories \cite{Ferrari:2000ep,Konoplya:2001ji,Brito:2013wya,Brito:2013yxa,Babichev:2015zub,Molina:2010fb,Pani:2009wy,Blazquez-Salcedo:2016enn,Blazquez-Salcedo:2017txk,Brito:2018hjh,Pierini:2022eim,Dias:2015wqa,Wagle:2021tam,Cano:2020cao,Cano:2021myl,Tattersall:2018nve}, of which even fewer theories have QNMs computed at a beyond-leading order in BH spins \cite{Dias:2015wqa,Pierini:2022eim}, we consider both publicly available modified spectra and phenomenologically modified spectra. 

The spectra are chosen to encompass a variety of modifications to stress-test the ability of spectroscopy to distinguish them from a GR Kerr spectrum. We don't concern ourselves with the physical plausibility of these modifications. Below we describe the modifications to the GR Kerr BH spectrum used in this study:
\begin{itemize}
    \item EdGB: The Einstein-dilaton-Gauss-Bonnet theory \cite{Mignemi:1992nt,Kanti:1995vq} is a modified theory of gravity that introduces a dilaton scalar field that is non-minimally coupled to higher orders of the curvature, specifically to the Gauss-Bonnet invariant. The BHs in EdGB have a scalar hair as they are endowed with a monopole scalar charge. However, this charge is not an independent parameter but it is a ``secondary hair" \cite{Herdeiro:2015waa}, i.e., it is completely determined by the mass and spin of the BH and by the coupling constants of the theory. The QNMs of EdGB BHs at the next-to-leading order in the spin are derived in \cite{Pierini:2022eim}. The numerical approximations in \cite{Pierini:2022eim} restrict the validity of the spectrum to $\zeta_{\rm EdGB}\lesssim0.4$, where $\zeta_{\rm EdGB}=\alpha_{\rm EdGB}/M_f^2$ and here $\alpha_{\rm EdGB}$ is the coupling constant of the theory. Further, the final spins $\chi_f\gtrsim0.3$ can be potentially outside the range of validity of the $\mathcal{O}(\chi_f^2)$ expansion. To mitigate these effects, we consider the Padé resummed version of the spectrum provided in \cite{Pierini:2022eim}. Note that the EdGB QNMs break isospectrality \cite{chandrasekhar1998mathematical} between axial and polar sectors due to the non-minimal coupling of the scalar field. In this work, we choose to not include the axial sector in the spectrum and model it as given by the polar sector only.
    
    \item Kerr-Newman: The Kerr-Newman spectrum for GW perturbation is derived in \cite{Pani:2013ija,Pani:2013wsa} at first order in the final spin expansion (c.f., \cite{Mark:2014aja} for a perturbative expansion in the electric charge). There is strong numerical evidence that the Kerr-Newman spectrum is isospectral, which is confirmed by the full non-perturbative analysis in \cite{Dias:2015wqa}. Therefore, unlike in the EdGB, there is no ambiguity in choosing polar or axial sectors in its modified spectrum. Note that a Kerr-Newman BH becomes extremal at charge-to-mass ratio $Q=(1-\chi_f^2)^{1/2}$ but we only consider values of $Q$ away from this limit.
    
   \item Horndeski: The Horndeski action gives a general scalar-tensor gravity with second order equations of motion \cite{horndeski1974second}. The Horndeski field equations admit standard GR BH solutions under various conditions \cite{Motohashi:2018wdq}. Linear perturbations around slowly rotating Kerr BHs were studied in \cite{Tattersall:2018nve} for the sub-class of Horndeski theories in which GW propagates at the speed of light. They show that the equations are reduced to a massive scalar perturbation with an effective mass parameter $\mu$. Although the spectrum does not correspond to perturbations in GW sector, in principle, it can be sourced by the GW sector, and we expect imprints of these frequencies in the GW signals \cite{Evstafyeva:2022rve}. In this work, we only look at the QNMs in scalar sector presented in Eq.s~(34)-(35) of \cite{Tattersall:2018nve}. Note that this spectrum reduces to the \textit{scalar} perturbations of the Kerr BH in the limit $\mu\to0$; therefore, to augment our battery of modifications, we linearly re-scale it to recover the \textit{gravitational} GR Kerr BH spectrum in the limit $\mu\to0$ and promote it as yet another modified spectrum. We remind that for this work, we are interested in studying the performance of BH spectroscopy to distinguish a non-Kerr GR spectrum and do not aim to put bounds on any given modified theory/spectrum in particular.

    \item dCS: In dynamical Chern-Simons (dCS) theory \cite{Alexander:2009tp}, a scalar field is non-minimally coupled to the higher-curvature Pontryagin invariant, resulting in a breakdown of parity symmetry. Rotating BHs in dCS have a secondary hair in the form of a monopole scalar charge \cite{Yagi:2012ya}. The QNM spectrum of dCS BHs were computed in \cite{Wagle:2021tam} at the leading order in the spin and at second order in the non-minimal coupling constant of the scalar field $\alpha_{\rm dCS}$ (see also \cite{Srivastava:2021imr}). In the following we will use the dimensionless coupling $\zeta_{\rm dCS}=\alpha^2_{\rm dCS}/M_f^4$. Here, we are extrapolating the spectrum in \cite{Wagle:2021tam} beyond the small spin approximation but this is not a critical concern for our study. Also, the non-minimal coupling of the scalar field breaks isospectrality and therefore, similar to EdGB we consider the polar sector of the dCS spectrum.
    \item Delta: We generate an ad-hoc phenomenological spectrum by modifying the frequencies of all modes by a constant relative shift, $f_{lmn}=f_{lmn}^{\rm Kerr}(1+\Delta)$. We also choose to leave all damping times unchanged.
    \item Delta220: We modify only the the frequency of the dominant mode $f_{220}=f_{220}^{\rm Kerr}(1+\Delta_{220})$ and leave all other mode frequencies and damping times unchanged.
 
    \end{itemize}

The above scenarios are distinct modifications to the GR Kerr BH spectra where the no-hair hypothesis can be violated. While Kerr-Newman BHs deviate from the Kerr background due to the presence of a ``primary hair" (the electric charge), in the case of EdGB and dCS BHs the background possesses a ``secondary hair" (the monopole scalar charge). Further in the Horndeski BHs we consider here, the background coincides with Kerr. Moreover, note that in all the theories here, the deviations from GR appear also at the level of the field equations.

The Delta and Delta220 are simplistic ad-hoc modification schemes. We study them as they are easy to implement and to interpret the performance of BH spectroscopy and as a benchmark. We opt for modifications in the frequencies as deviations in the damping times are more difficult to constrain. Note that, in a realistic scenario, we typically expect all or at least a subset of frequencies and damping times to be modified and it is unlikely that all modes are modified by the same amount. Nonetheless, as shown in Fig.~\ref{fig:qgr}, they behave as simpler representatives of the more complex spectra considered in this work.

For the EdGB, Kerr-Newman, and dCS spectra, we opt to impose consistency with the Kerr BH spectrum by linear rescaling, similar to the Horndeski case described above. If we take the limit of $\zeta_{\rm EdGB}\to0$ for the EdGB spectrum in \cite{Pierini:2022eim}, we do not recover the GR Kerr BH QNMs (which would be the case if the EdGB QNMs could be computed non-perturbatively) because the spectrum is derived at $\mathcal{O}(\chi_f^2)$. Additionally, each spectrum is derived within its own set of approximations, and therefore, they return a different approximation to the GR Kerr BH spectrum in the limit of the vanishing deviation parameters. 

We provide a procedure for imposing consistency across the spectra by redefining the spectra. We illustrate our procedure on EdGB below -- 
\begin{equation}
\begin{split}
    \label{eq:rescale:1}
    &f_{lmn}^{\rm EdGB}(M_f,\chi_f,\zeta_{\rm EdGB})\\
    &=f_{lmn}^{\rm Kerr}(M_f,\chi_f)\left(\frac{
    \hat{f}_{lmn}^{\rm EdGB}(M_f,\chi_f,\zeta_{\rm EdGB})}{
    \hat{f}_{lmn}^{\rm EdGB}(M_f,\chi_f,0)}\right)
\end{split}
\end{equation}

We enforce the GR Kerr BH spectrum in the limit $\zeta_{\rm EdGB}\to0$. Here, the hat denotes the expression of EdGB spectrum presented in \cite{Pierini:2022eim}. Then, we defined the relative shifts as -- 
\begin{equation}
    \label{eq:rescale:2}
    \delta f_{lmn} = \frac{\hat{f}_{lmn}^{\rm EdGB}(M_f,\chi_f,\zeta_{\rm EdGB})}{\hat{f}_{lmn}^{\rm EdGB}(M_f,\chi_f,0)}-1
\end{equation}
and parametrized
\begin{equation}
    \label{eq:rescale:3}
    f_{lmn}^{\rm EdGB}=f_{lmn}^{\rm Kerr}(1+\delta f_{lmn})\,.
\end{equation}
The definition \eqref{eq:rescale:2} preserves the values of the shifts given by the traditional parametrization of the spectrum. We also repeat the same procedure for the damping times. The procedure is extended to the other spectra considered above, by replacing $\zeta_{\rm EdGB}$ with $Q,\mu,\zeta_{\rm dCS},\Delta$ and $\Delta_{220}$ respectively.

Here, we re-emphasis that this procedure is only necessary because the QNM spectra in these modified theories are calculated perturbatively to a limited order in the final spin of the BH. In the absence of the exact spectra, we use Eq.~\eqref{eq:rescale:3} as a fiducial definition for all the modified spectra considered here.

\begin{figure*}[p]
    \centering
    \includegraphics[width=0.43\textwidth]{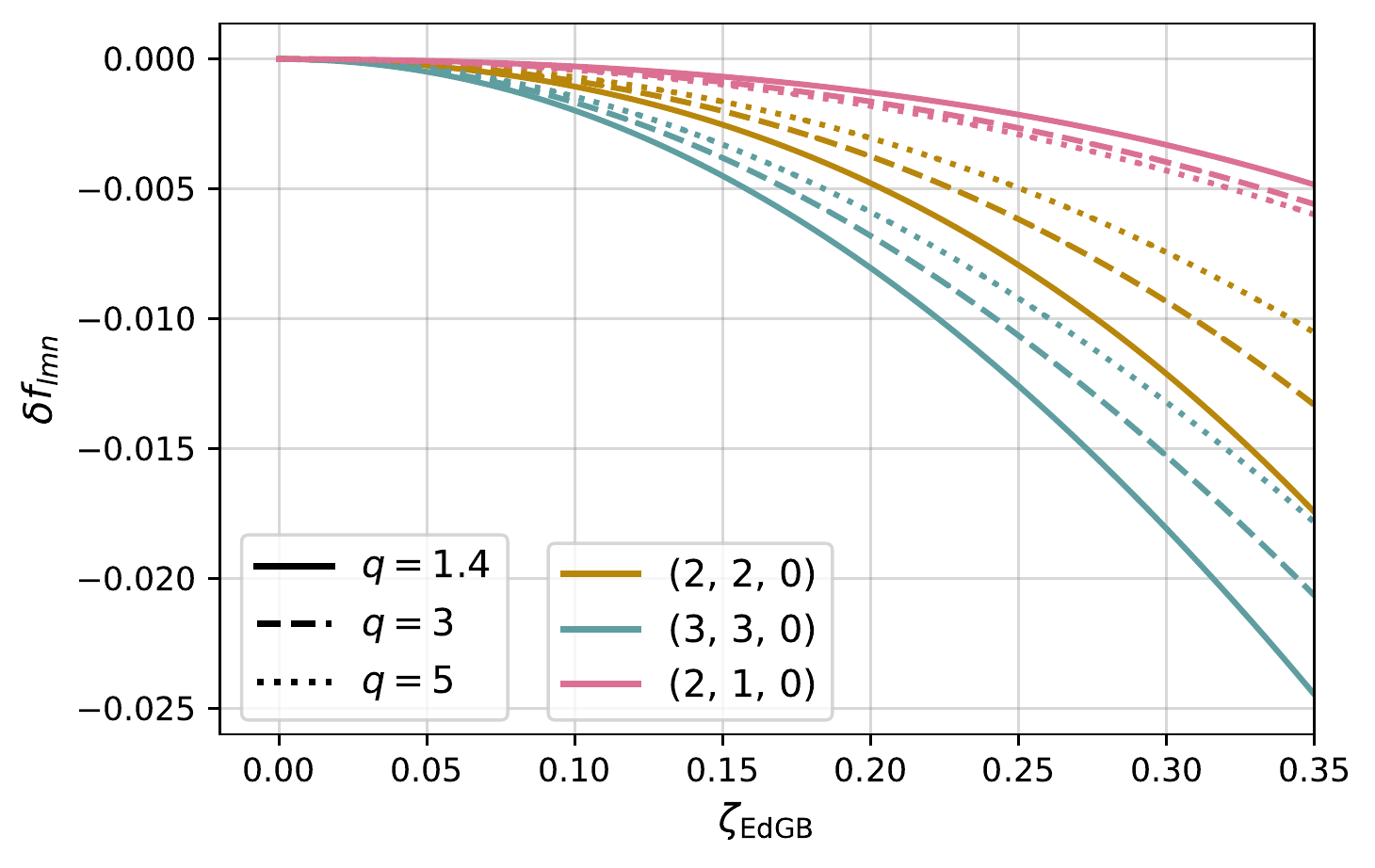}
    \includegraphics[width=0.43\textwidth]{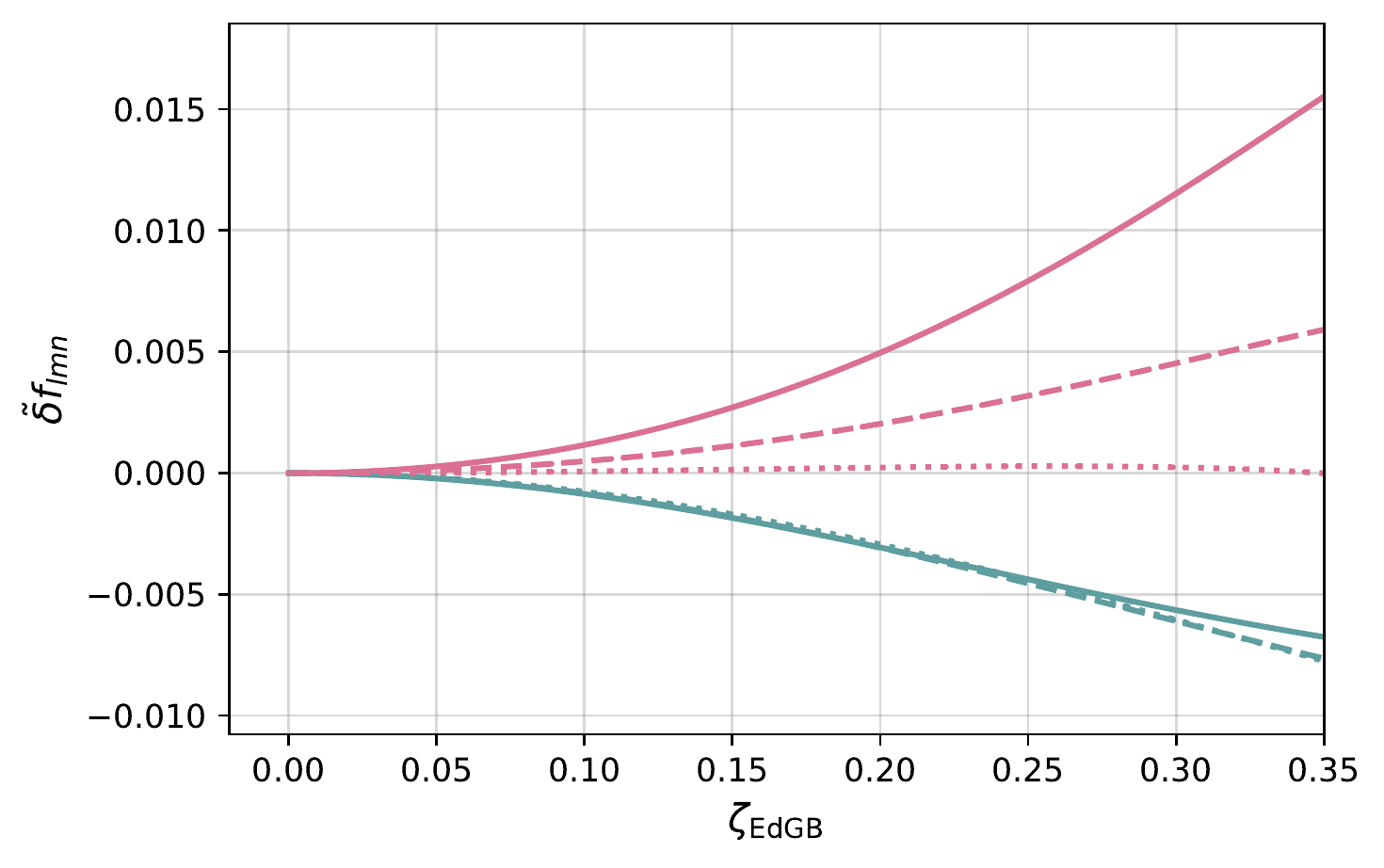}\\
    \includegraphics[width=0.43\textwidth]{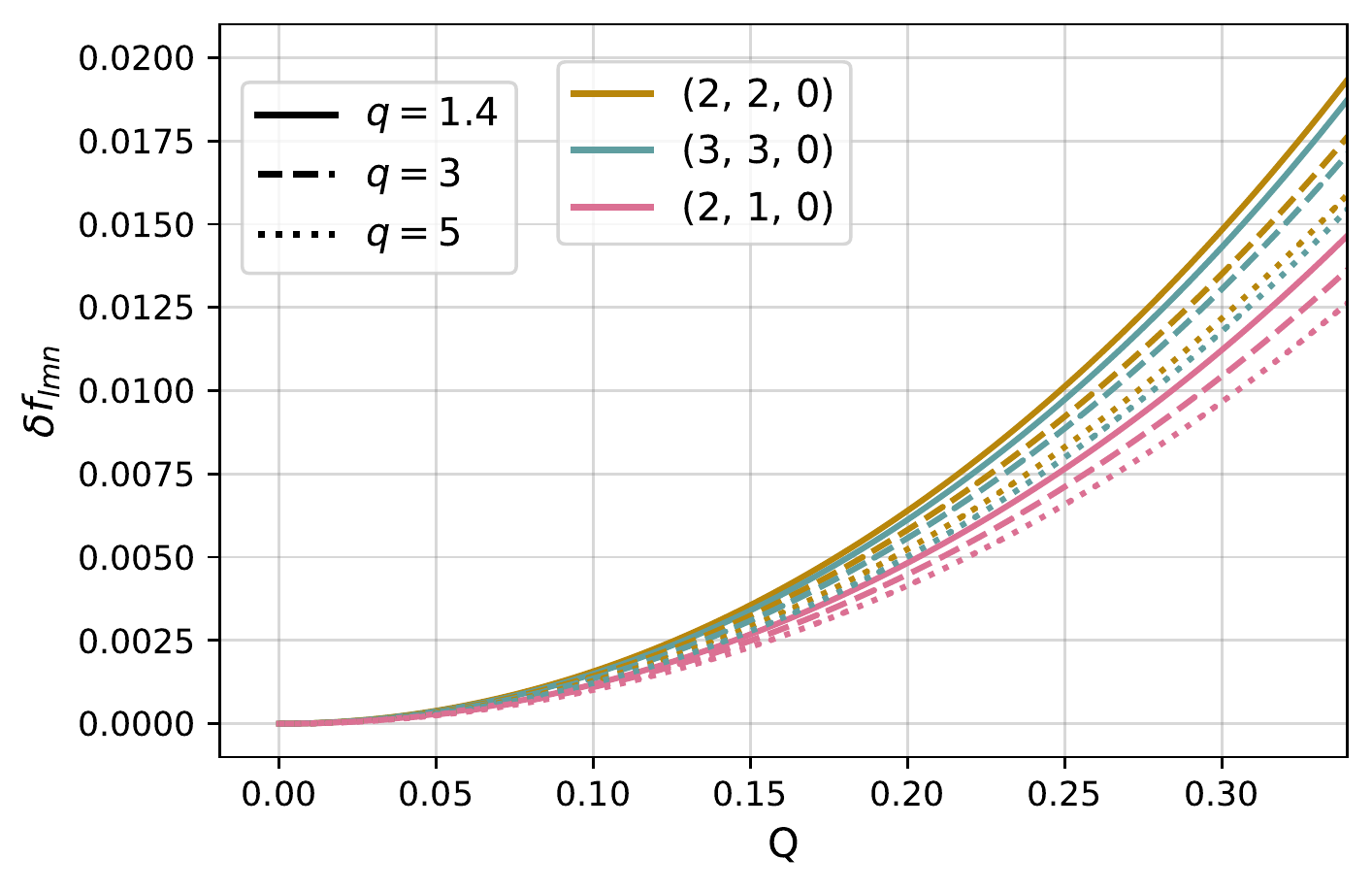}
    \includegraphics[width=0.43\textwidth]{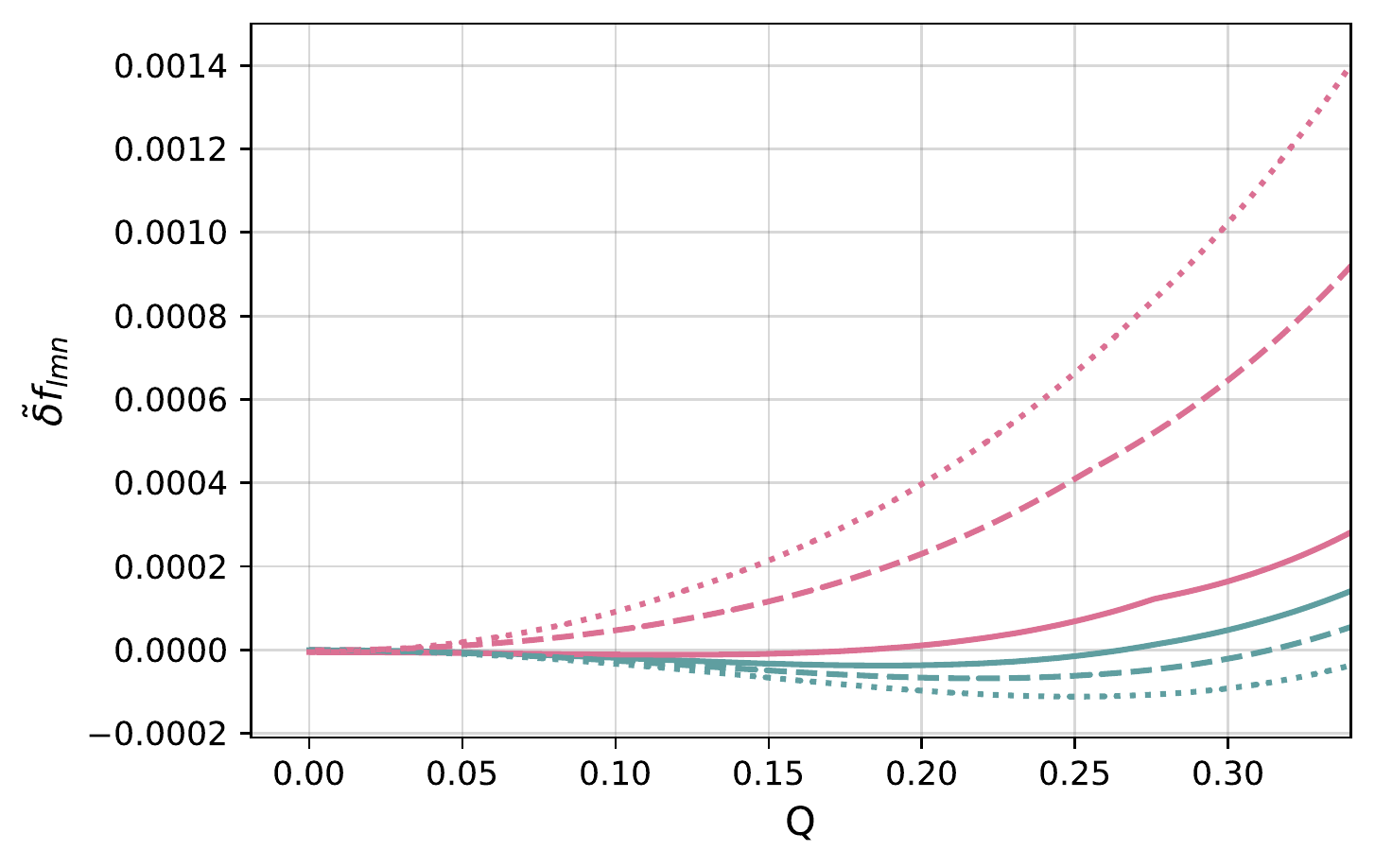}\\
    \includegraphics[width=0.43\textwidth]{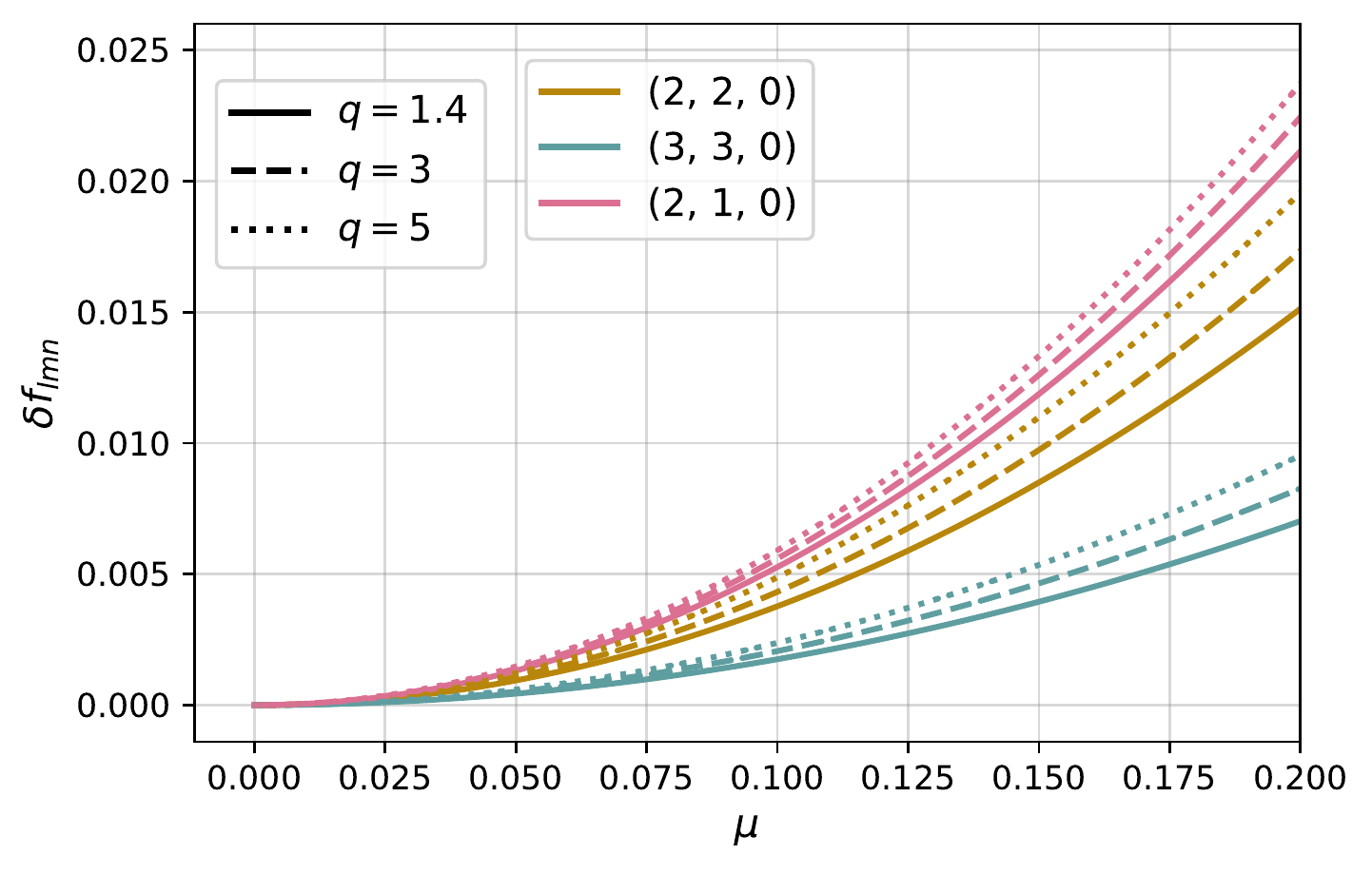}
    \includegraphics[width=0.43\textwidth]{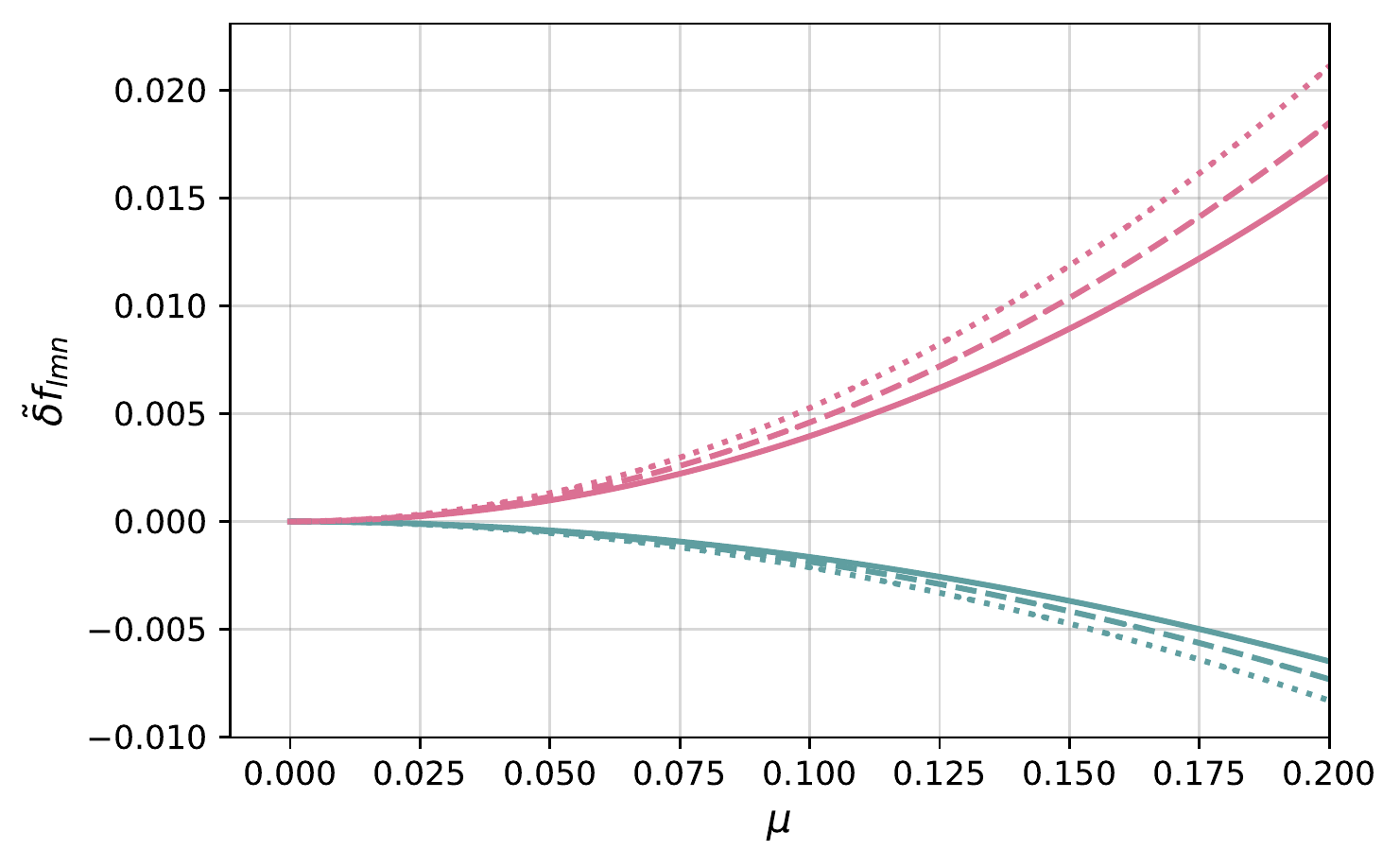}\\
    \includegraphics[width=0.43\textwidth]{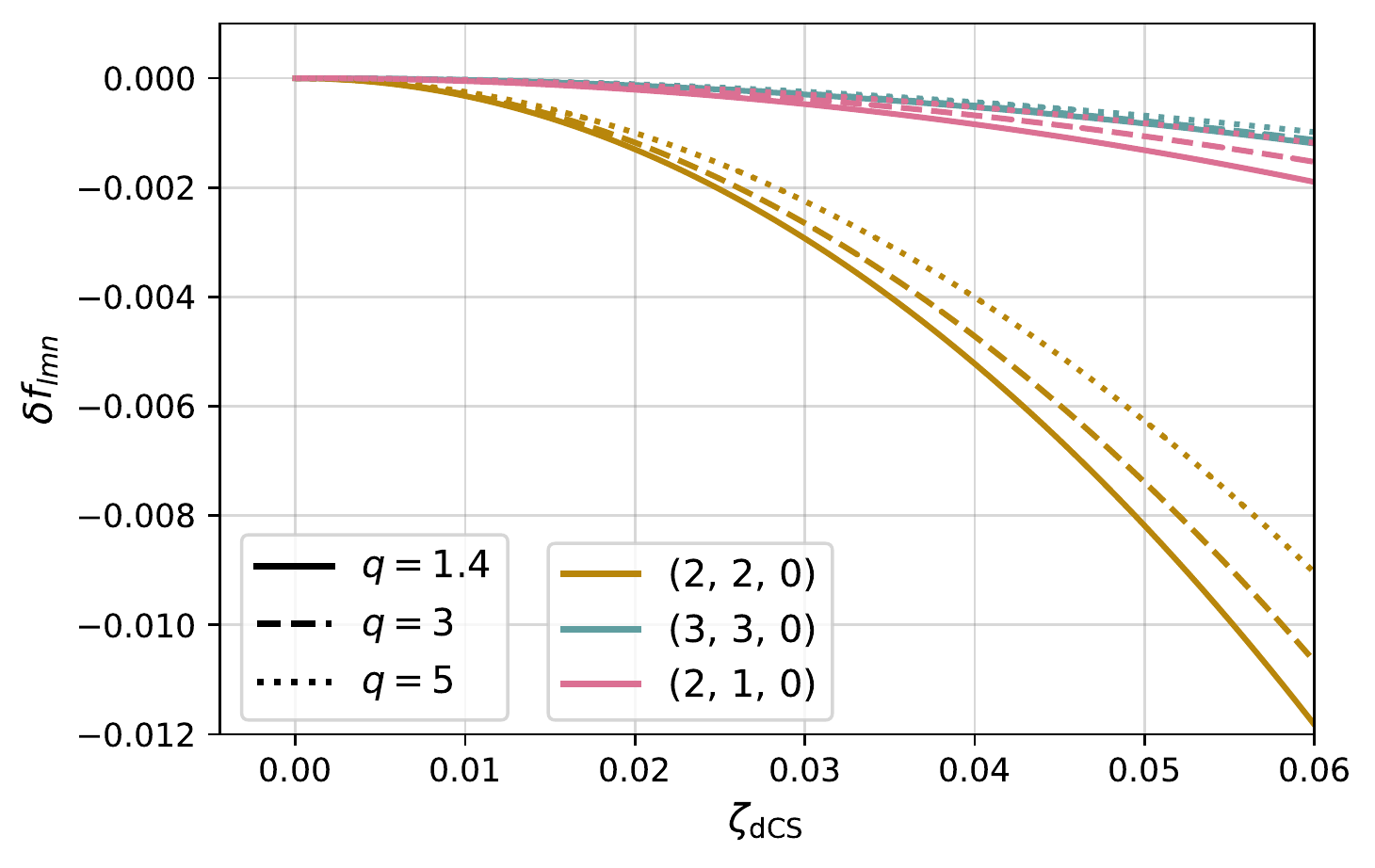}
    \includegraphics[width=0.43\textwidth]{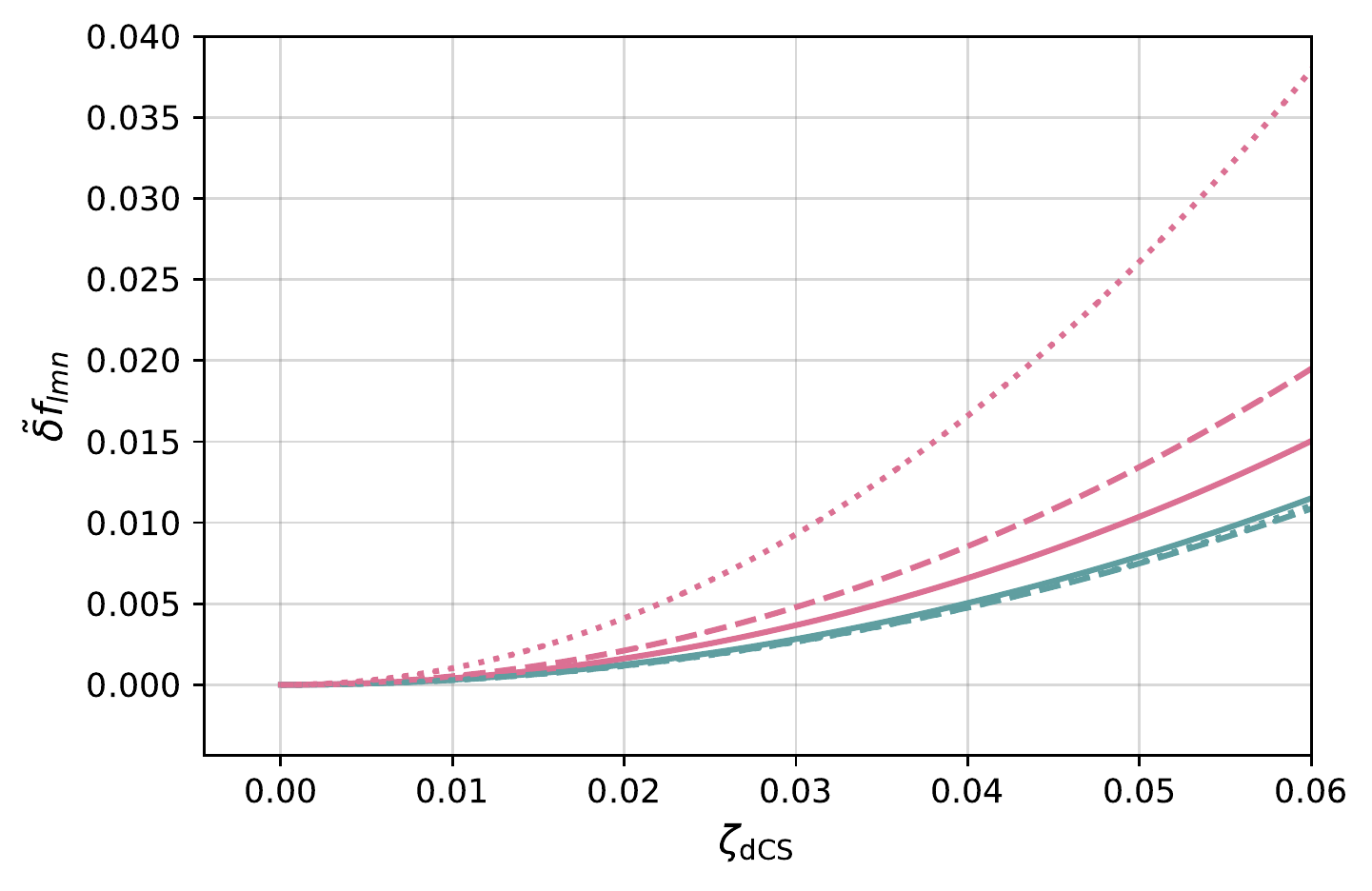}\\
    \includegraphics[width=0.43\textwidth]{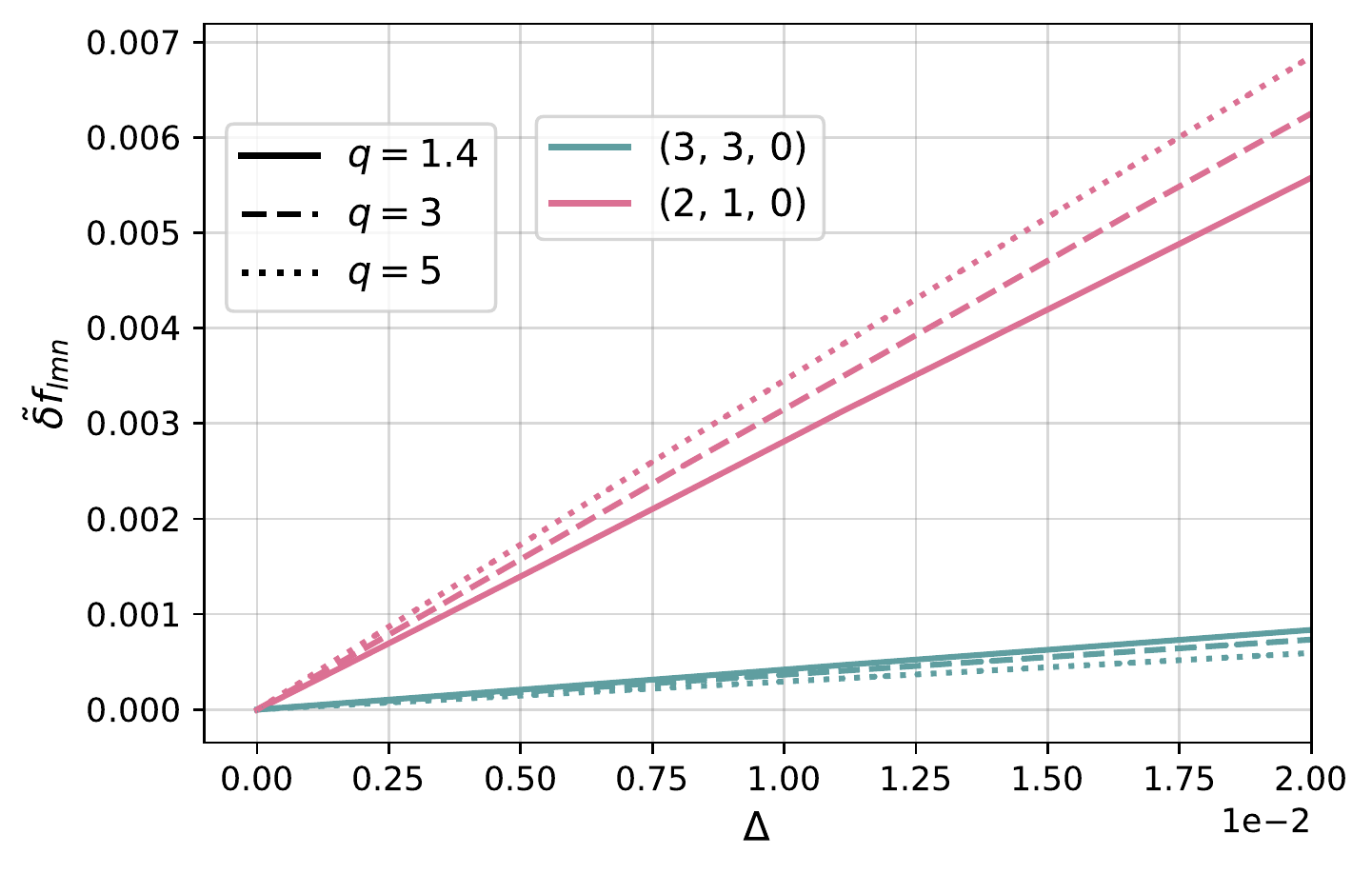}
    \includegraphics[width=0.43\textwidth]{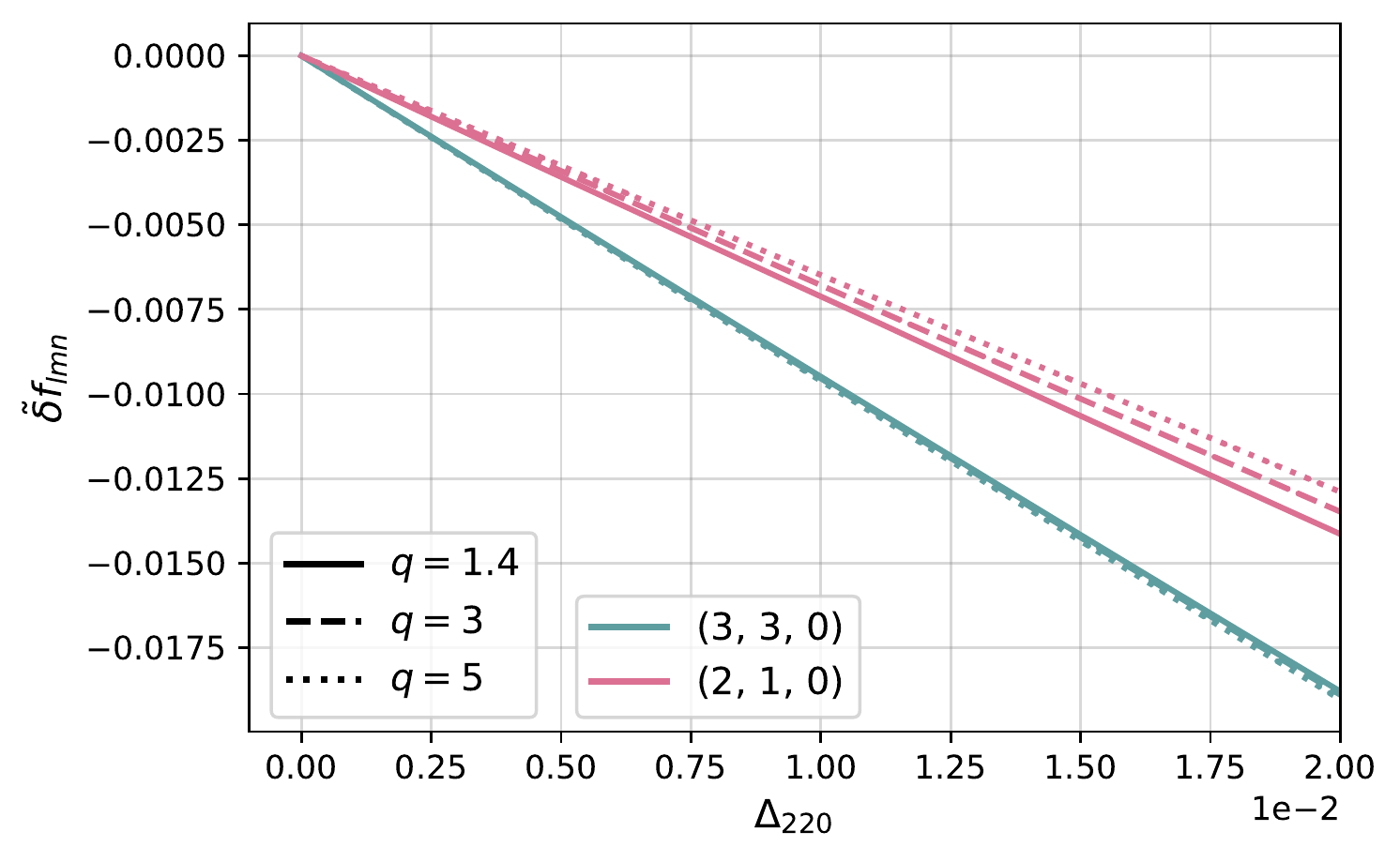}
    \caption{Rows 1-4: Relative deviations $\delta f_{lmn}$ (left) and relative effective deviations $\tilde{\delta} f_{lmn}$ (right) from the GR Kerr BH spectrum, as induced by the different modified spectra discusses in Sec.~\ref{sec:spectra}: EdGB, Kerr-Newman, Horndeski and dCS, respectively. Last row: Relative effective deviations $\tilde{\delta} f_{lmn}$ induces by the Delta spectrum (left) and the Delta220 spectrum (right). }
    \label{fig:spectra}
\end{figure*}

Finally, in Fig.~\ref{fig:spectra} we plot the relative deviations $\delta f_{lmn}$ and effective deviations $\tilde{\delta} f_{lmn}$ in the QNM frequencies for the modified spectra listed above. We remind that it is the effective deviations $\tilde{\delta} f_{lmn}$ that are measured when performing BH spectroscopy. Interestingly, we find that for some spectra even when the actual spectrum has deviations at a percent level, the measurable effective spectrum deviates from GR Kerr BH at a much smaller sub-percent level. This is particularly evident for the Delta spectrum where all the true frequencies are shifted by the same amount but the measurable deviations turn out to be much smaller. We also see this in the Kerr-Newman spectrum.
\begin{table*}[t]
    \centering
    \begin{tabular}{|c|c|c|c|c|c|c|c|}
         \hline
         Spectrum & $\alpha_{0.01}$ & $\tilde{M}_f$ ($M_\odot$)& $\tilde{\chi}_f$ & $\tilde{\delta} f_{330}$ & $\tilde{\delta} f_{210}$
         & $\tilde{\delta} \tau_{330}$ & $\tilde{\delta} \tau_{210}$\\
         \hline
         \multicolumn{6}{c}{$q=1.4\,,\quad\chi_f=0.67$}\\
         \hline
         EdGB & $0.28$ & $71.10$ & $0.68$ & $-5.1\times10^{-3}$ & $9.7\times10^{-3}$ & $-1.2\times10^{-2}$ & $-1.6\times10^{-2}$\\
         Kerr-Newman & $0.25$ & $69.71$ & $0.68$ & $-1.6\times10^{-5}$ & $6.6\times10^{-5}$ & $3.9\times10^{-4}$ & $1.9\times10^{-4}$\\
         Horndeski & $0.16$ & $70.38$ & $0.69$ & $-4.3\times10^{-3}$ & $1.1\times10^{-2}$ & $-8.3\times10^{-3}$ & $9.9\times10^{-3}$\\
         dCS & $0.055$ & $71.40$ & $0.68$ & $9.7\times10^{-3}$ & $1.3\times10^{-2}$ & $-3.7\times10^{-2}$ & $1.2\times10^{-2}$\\
         Delta & $0.01$ & $69.76$ & $0.68$ & $4.2\times10^{-4}$ & $2.8\times10^{-3}$ & $-4.7\times10^{-4}$ & $6.4\times10^{-4}$\\
         Delta220 & $0.01$ & $69.76$ & $0.68$ & $-9.5\times10^{-3}$ & $-7.1\times10^{-3}$ &
         $-4.7\times10^{-4}$ & $6.4\times10^{-4}$\\
         \hline
         \multicolumn{6}{c}{$q=3\,,\quad\chi_f=0.54$}\\
         \hline
         EdGB & $0.31$ & $70.54$ & $0.54$ & $-6.4\times10^{-3}$& $4.8\times10^{-3}$
         & $-7.6\times10^{-3}$ & $-9.3\times10^{-3}$\\
         Kerr-Newman & $0.26$ & $69.75$ & $0.55$ & $-5.7\times10^{-5}$ & $4.5\times10^{-4}$ &
         $3.9\times10^{-4}$ & $-2.7\times10^{-5}$\\
z         Horndeski & $0.15$ & $70.57$ & $0.57$ & $-4.3\times10^{-3}$ & $1.1\times10^{-2}$ &
         $-8.6\times10^{-3}$ & $6.9\times10^{-3}$\\
         dCS & $0.058$ & $72.30$ & $0.58$ & $1.0\times10^{-2}$ & $1.8\times10^{-2}$ &
         $-5.7\times10^{-2}$ & $-3.1\times10^{-2}$
         \\
         Delta & $0.01$ & $69.82$ & $0.55$ & $3.6\times10^{-4}$ & $3.1\times10^{-3}$ &
         $-4.5\times10^{-4}$ & $4.0\times10^{-4}$\\
         Delta220 & $0.01$ & $69.82$ & $0.55$ & $-9.5\times10^{-3}$ & $-6.8\times10^{-3}$
         & $-4.5\times10^{-4}$ & $4.0\times10^{-4}$\\
         \hline
         \multicolumn{6}{c}{$q=5\,,\quad\chi_f=0.42$}\\
         \hline
         EdGB & $0.34$ & $70.01$ & $0.40$ & $-7.5\times10^{-3}$ & $5.2\times10^{-5}$
         & $-4.7\times10^{-3}$ & $-4.7\times10^{-3}$\\
         Kerr-Newman & $0.27$ & $69.77$ & $0.43$ & $-1.1\times10^{-4}$ & $8.2\times10^{-4}$ &
         $4.7\times10^{-4}$ & $-9.3\times10^{-5}$\\
         Horndeski & $0.14$ & $70.71$ & $0.46$ & $-4.3\times10^{-3}$ & $1.1\times10^{-2}$ &
         $-8.6\times10^{-3}$ & $4.8\times10^{-3}$\\
         dCS & $0.063$ & $76.26$ & $0.55$ & $1.2\times10^{-2}$ & $4.2\times10^{-2}$ &
         $-1.3\times10^{-1}$ & $-9.6\times10^{-2}$\\
         Delta & $0.01$ & $69.86$ & $0.43$ & $2.9\times10^{-4}$ & $3.5\times10^{-3}$ &
         $-3.7\times10^{-4}$ & $2.8\times10^{-4}$
         \\
         Delta220 & $0.01$ & $69.86$ & $0.43$ & $-9.6\times10^{-3}$ & $-6.5\times10^{-3}$
         & $-3.7\times10^{-4}$ & $2.8\times10^{-4}$\\
         \hline
    \end{tabular}
    \caption{Values $\alpha_{0.01}$ of the deviation parameter $\alpha$ inducing a $1\%$ shift in $f_{220}$, alongside the effective measurable final mass and spin $\{\tilde{M}_f,\tilde{\chi}_f\}$ as well as the effective shifts $\{\tilde{\delta} f_{lmn},\tilde{\delta}\tau_{lmn}\}$ of the subdominant modes, for the different spectra considered in this work. Here $\alpha$ is a collective name for the additional theory parameter.  It represents $\{\zeta_{\rm EdGB},Q,\mu,\zeta_{\rm dCS},\Delta,\Delta_{220}\}$ depending on the spectrum.}
    \label{tab:tilde}
\end{table*}
 For each spectrum and mass ratio, we quantify these differences in Tab.~\ref{tab:tilde}. We present the values of the deviation parameters $\alpha$ such that the deviation in the dominant mode frequency is at $1 \%$ level i.e., $|\delta f_{220}|=0.01$. In Sec.~\ref{sec:results}, using the values in Tab.~\ref{tab:tilde}, we  study the ability of BH spectroscopy to constrain $\tilde{\delta}f_{lmn}$ away from zero i.e., exclude the GR Kerr BH spectrum. 

\section{Methods and implementations}
\label{sec:methods}
We construct ringdowns that comprise of  $(l, m, n) \in \{(2,2,0), (3,3,0), (2,1,0) \}$ modes with the modified QNM spectra described in the previous Section \footnote{We do not consider overtones for simplicity as their measurablity and physical interpretation is still debated even when assuming that the underlying theory of gravity is GR \cite{Bhagwat:2019dtm,Cotesta:2022pci,Forteza:2020cve,Dhani:2020nik, Isi:2022mhy, Cheung:2022rbm,Mitman:2022qdl}.}. We remind that the coupling parameters of the theory or the extra charges that modify the QNM spectra are chosen such that the dominant-mode frequency $f_{220}$ differs from the GR Kerr value $f_{220}^{\rm Kerr}$ by $1 \%$. This is a heuristic choice but studying the effect of $1 \%$ deviation in $f_{220}$ is a reasonable goal from the data analysis perspective as the next-generation detectors is expected to measure $f_{220}$ with sub-percent accuracy \cite{Maselli:2017kvl,Shi:2019hqa,Bhagwat:2021kwv}. For Delta and Delta220 modifications, the sign of the deviation is chosen to be positive i.e., fractional frequency shift of $ + 0.01$. 

We fix the final mass $M_f = 70 M_\odot$ in the detector frame. To compute $\rho_{\rm RD}$, we consider events with extrinsic parameters listed in Tab.~\ref{tab:extrinsic:params}. These choices are compatible with the GW150914 event, with the only exception of the inclination angle; the posteriors distribution for the inclination angle of GW150914 prefers a face-on/face-off orientation for which the subdominant modes excitation is suppressed significantly. Instead, we use a more optimal inclination angle for BH spectroscopy and set $\iota=\pi/3$. For each modified QNM scenario, we study mass ratio $q \in \{1.4, 3, 5 \}$.  The final spin of the remnant and the mode excitation amplitudes $\mathcal{A}_{lmn}$ are computed from the $q$ using the numerical fits provided in \cite{Hofmann:2016yih} and \cite{Forteza:2022tgq}, respectively, by assuming non-spinning progenitors, while the absolute amplitude scale $\mathcal{A}_{220}$ is set by Eq.~(5) in \cite{Gossan:2011ha}\footnote{Here we emphasize that depending on the modified theory, these assumptions may not hold to differing extents. Amplitudes are governed dominantly by the plunge-merger dynamics and we typically need a fully numerical simulation to infer them. However, we do not have numerical simulations for most of the theories considered here; thus, we do not have knowledge of the amplitudes of mode excitations in these theories. Moreover, there can be differences in the energy and momentum radiated and therefore, in the expressions for the remnant mass and spin. Since we are only interested in providing order of magnitude estimates, we use the GR amplitudes and remnant properties as an approximate proxy for practicality.}. In particular, $\chi_f\approx\{0.67,0.54,0.52\}$ for the three values of $q$ chosen here. The phases $\phi_{lmn}$ do not significantly affect the recovery of the QNM frequencies and damping times (c.f., \cite{Bhagwat:2019dtm,Berti:2005ys}) and therefore for this study, we set it to $0$.

The ringdown waveform is modeled as $h=h_++ih_\times$, with
\begin{subequations}
    \label{eq:waveform}
    \begin{align}
        &h_+=\sum_{l,m>0,n}\mathcal{A}_{lmn}~{_{-2}}\tensor{Y}{_{lm,+}}{}(\iota)e^{-t/\tau_{lmn}}\cos\Phi_{lmn}\,,
    \label{eq:waveform:a}\\
        &h_\times=\sum_{l,m>0,n}\mathcal{A}_{lmn}~{_{-2}}\tensor{Y}{_{lm,\times}}{}(\iota)e^{-t/\tau_{lmn}}\sin\Phi_{lmn}\,,
    \label{eq:waveform:b}
    \end{align}
\end{subequations}
where $\Phi_{lmn}=2\pi f_{lmn}t+\phi_{lmn}$, $\mathcal{A}_{lmn}$ and $\phi_{lmn}$ are the (real) excitation amplitudes and phases of the modes, and the plus and cross spherical harmonics are defined by\footnote{A global phase in the definition of the spherical harmonics can be reabsorbed in the definition of the phases $\phi_{lmn}$ of the ringdown modes.}
\begin{subequations}
    \label{eq:harmonics}
    \begin{align}
    &{_{-2}}\tensor{Y}{_{lm,+}}{}(\iota)={_{-2}}\tensor{Y}{_{lm}}(\iota,0)+(-1)^l{_{-2}}\tensor{Y}{_{lm}}(\iota,0)\,,
    \label{eq:harmonics:a}\\
    &{_{-2}}\tensor{Y}{_{lm,+}}{}(\iota)={_{-2}}\tensor{Y}{_{lm}}(\iota,0)-(-1)^l{_{-2}}\tensor{Y}{_{lm}}(\iota,0)\,.
    \label{eq:harmonics:b}
    \end{align}
\end{subequations}

In writing the expressions \eqref{eq:waveform:a}-\eqref{eq:waveform:b}, we assume a non-precessing quasi-circular binary progenitor. For these systems equatorial reflection symmetries gives $\mathcal{A}_{l-mn}e^{i\phi_{l-mn}}=(-1)^l\mathcal{A}_{lmn}e^{-i\phi_{lmn}}$ and we simultaneously sum over $\pm m$ using the symmetry relation $\omega_{l-mn}=-\omega^*_{lmn}$\footnote{We use $\omega_{lmn}$ to denote the prograde modes of the spectrum. The QNM solutions also contain a set of retrograde modes, but in the numerical simulations it is seen that these modes are not excited significantly \cite{London:2014cma}. See also \cite{Dhani:2020nik, Finch:2021iip} on the role of retrograde modes in describing the ringdown waveform.}. We assume a circularly polarized ringdown as the numerical simulations favour it  (c.f., \cite{Borhanian:2019kxt} for a details). We also use spherical harmonics instead of the more natural spheroidal harmonics basis function for ringdown; this is a fairly standard approximation which is known to introduces substantial errors only in high spin limits \cite{Berti:2005gp,Berti:2014fga}.

To predict the statistical uncertainty in the measurement of $\tilde{\delta} f_{lmn}$, we employ a Fisher matrix formalism following our work in \cite{Bhagwat:2021kwv}. We use the power spectral density (PSD) of the next-generation ground-based detector -- the Einstein Telescope, assuming a triangular geometry, located in Sardinia \cite{Gossan:2021eqe} and operating at the ET-D sensitivity \cite{Hild:2010id}. However, note that all the qualitative statements and ball-park numbers reported in this study should hold for any detector for a given $\rho_{\rm RD}$. Our study is not heavily influenced by the choice of the detector as long as the modifications in the spectra are chosen to have $1\%$ deviation of $f_{220}$. Note that while the results in this study will be approximately similar for the next space-based detector LISA, LISA will be sensitive to supermassive binary BH signals and therefore, the theory-specific parameters i.e., coupling constants or extra charges, that lead to $1 \%$ deviation of $f_{220}$ will be different from the choice made here. However, we expect theory-parameters similar to the choice made here for the Cosmic Explorer, as it is sensitive to stellar mass ranges similar to the Einstein Telescope \cite{Berti:2016lat}.

For computing the Fisher covariance matrix, we parametrize the waveform as Eq.s~\eqref{eq:reparam:1:a}-\eqref{eq:reparam:1:b} and \eqref{eq:reparam:2:a}-\eqref{eq:reparam:2:b}, i.e., using effective parameters  $\{\tilde{M}_f,\tilde{\chi}_f,\tilde{\delta} f_{lmn},\tilde{\delta}\tau_{lmn}\}$ instead of the traditionally used $\{f_{lmn},\tau_{lmn}\}$. Next, for a choice of theory parameters that produces a modified spectra such that $f_{220}$ deviates by $1\%$ w.r.t.~GR Kerr BH, we map the true values $\{M_f,\chi_f\}$ to the measured values $\{\tilde{M}_f,\tilde{\chi}_f\}$ using Eq.s~\eqref{eq:invert} and \eqref{eq:reparam:3:a}-\eqref{eq:reparam:3:b}. Note that once we compute the values of $\{\tilde{M}_f,\tilde{\chi}_f\}$, deviation parameters for all the subdominant modes get fixed. We construct the covariance matrix for a set of $4N$ parameters, where $N$ is the number of modes as --
\begin{equation}
    \label{eq:params}
    \bm{\theta}=\{\tilde{M}_f,\tilde{\chi}_f,\tilde{\delta} f_{lmn},\tilde{\delta}\tau_{lmn},\log_{10}\mathcal{A}_{lmn},\phi_{lmn}\}\,.
\end{equation}
For the dominant mode, we estimate $\{\tilde{M}_f,\tilde{\chi}_f\}$, and not $\{\tilde{\delta} f_{220}, \tilde{\delta} \tau_{220}\}$.

Next, we quantify the departure of the modified QNM spectra from GR Kerr QNM spectra using $Q_{\rm GR}$ as a measure, where $Q_{\rm GR}$ is defined as the quantile at which the posterior distribution of $\tilde{\delta}f_{lmn}$ excludes zero. Higher the value of  $Q_{\rm GR}$, more confidently we can exclude the null hypothesis that the ringdown contains QNMs corresponding to a GR Kerr BH. $Q_{\rm GR}$ can either be defined mode-wise over a marginalized 1-d probability distribution $P(\tilde{\delta} f_{lmn})$ or defined on a joint probability distribution of all the measured modes. Here, we use $Q_{\rm GR}$ defined on the 1-d posterior distributions corresponding to -- a) $\tilde{\delta} f_{330}$ : $Q_{\rm GR}^{330}$, b) $\tilde{\delta} f_{210}$ : $Q_{G\rm R}^{210}$ and c) on a joint posterior distribution of $\tilde{\delta} f_{330}$ and $\tilde{\delta} f_{210}$ : $Q_{\rm GR}^{\rm 2D}$.

In the Fisher matrix formalism, by construction, $P(\tilde{\delta}f_{lmn})$ is a Gaussian distribution centered at the true value of $\tilde{\delta} f_{lmn}$. From this, $Q^{lmn}_{\rm GR}$ can be easily computed as 
\begin{equation}
    \label{eq:quantile:1}
    Q^{lmn}_{\rm GR}={\rm erf}\left(p/\sqrt{2}\right)\,,
\end{equation}
where the $p$-value is given as
\begin{equation}
    \label{eq:quantile:2}
    p=\frac{|\tilde{\delta}f_{lmn}|}{\sigma}\,.
\end{equation}

Further, Eq.~\eqref{eq:quantile:2} can be generalized to a 2-dimensional case of the joint posterior $P(\tilde{\delta} f_{330},\tilde{\delta} f_{210})$. In the cases of 2-D posteriors --
\begin{equation}
    \label{eq:quantile:3}
    Q_{\rm GR}^{\rm 2D}=1-\exp\left(-\frac{1}{2}\vec{\mu}^{T}\cdot\mathbf{\Sigma}^{-1}\cdot\vec{\mu}\right)
\end{equation}
where $\vec{\mu}$ is the vector corresponding to the true values of $\{\tilde{\delta} f_{330},\tilde{\delta} f_{210}\}$ and $\mathbf{\Sigma}$ is the covariance matrix.
\section{Results}
\label{sec:results}
\begin{table}[t]
    \centering
    \begin{tabular}{|c |c |c |c |c |c|}
         \hline
         ra & dec & $\psi$ & $\iota$ & $t_{\rm GPS}$ & $d_L ({\rm Mpc})$ \\
         \hline
         $1.16$ & $-1.19$ & $1.12$ & $\pi/3$ & $1126259462.423$ & $403$\\
         \hline
    \end{tabular}
    \caption{Right ascension, declination, polarization angle, inclination angle, GPS time and luminosity distance for GW150914 that were used to compute the SNR and the Fisher covariance matrix.}
    \label{tab:extrinsic:params}
\end{table}

Before turning our attention to the modified spectra, we first look at the measurability of a GR Kerr BH spectrum with the next generation ground-based detectors. We used ET-D PSD to illustrate the expected orders of magnitudes and trends. Specifically, we consider ringdowns corresponding to $q \in \{1.4, 3, 5 \}$ with $M_{f} = 70 M_{\odot}$ and the extrinsic parameters displayed in Tab.~\ref{tab:extrinsic:params}. Tab.~\ref{tab:fisher} shows $\rho_{\rm RD}$ and the expected measurement uncertainties on the deviation parameters $\sigma(\tilde{\delta} f_{lmn})$ --- note that for a GR Kerr BH spectrum, $\tilde{\delta} f_{lmn} = \delta f_{lmn}$ and we use tilde here for consistency of notation. We report the normalized uncertainties $\kappa \equiv \rho_{\rm RD}\sigma$, where $\rho_{\rm RD}$ is the SNR in the ringdown and $\sigma$ is the expected statistical uncertainty in the parameter computed with a Fisher information matrix framework \footnote{Note, the relative excitation amplitudes $\mathcal{A}_{lmn}/\mathcal{A}_{220}$ of the subdominant modes increase monotonically with $q$ for non-spinning systems \cite{Kamaretsos:2011um,Kamaretsos:2012bs,Gossan:2011ha,Borhanian:2019kxt,Forteza:2022tgq}. However, $\rho_{\rm RD}$ decreases with $q$ \cite{Gossan:2011ha}. Therefore, while spectroscopy with high $q$ is favourable for fixed $\rho_{\rm RD}$, it need not be the case for fixed luminosity distance.}. Further, we confirm that the uncertainty in the measurement of the deviations in damping times $\sigma(\tilde{\delta}\tau_{lmn})$ is large; therefore, we concentrate on the subdominant mode frequencies.

\begin{table}[t]
    \centering
    \begin{tabular}{|c|c|c|c|c|c|c|c|}
         \hline
         $q$ & $\rho_{\rm RD}$ & \multicolumn{6}{c}{$\kappa=\rho_{\rm RD}\,\sigma$} \vline\\
         \hline
         \hline
         & & $\tilde{M}_f\,(M_\odot)$ & $\tilde{\chi}_f$ & $\tilde{\delta} f_{330}$ & $\tilde{\delta} f_{210}$ & $\tilde{\delta}\tau_{330}$ & $\tilde{\delta}\tau_{210}$\\
         \hline\hline
         1.4 & 87 & 184.63 & 3.68 & 1.56 & 7.78 & 22.43 & 46.14\\
         3 & 74 & 294.63 & 7.43 & 1.00 & 3.27 & 9.17 & 19.73\\
         5 & 57 & 436.37 & 13.08 & 1.27 & 3.38 & 8.80 & 22.26\\
         \hline
    \end{tabular}
    \caption{$\rho_{\rm RD}$ and uncertainties $\sigma$ over mass, spin and the deviations parameters for a Kerr BH spectrum, as measured by a triangular configuration of ET detector with ET-D PSD. The ringdown corresponds to a BH whose detector-frame final mass is $M_f=70 M_\odot$  and the  extrinsic system parameters are enlisted in Tab.~\ref{tab:extrinsic:params}.}
    \label{tab:fisher}
\end{table}

\begin{figure}[t]
    \centering
    \includegraphics[width=0.45\textwidth]{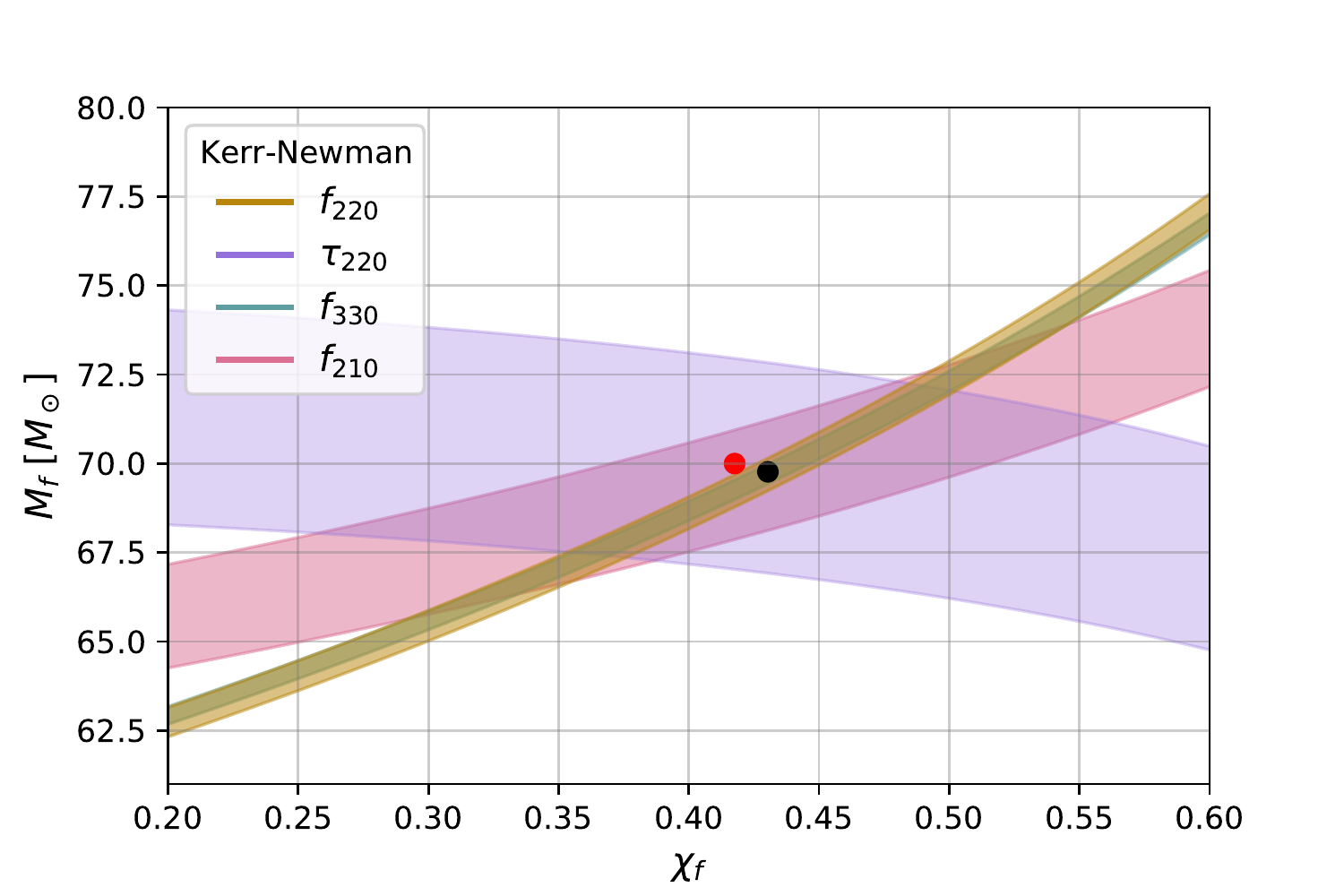}\\
    \includegraphics[width=0.45\textwidth]{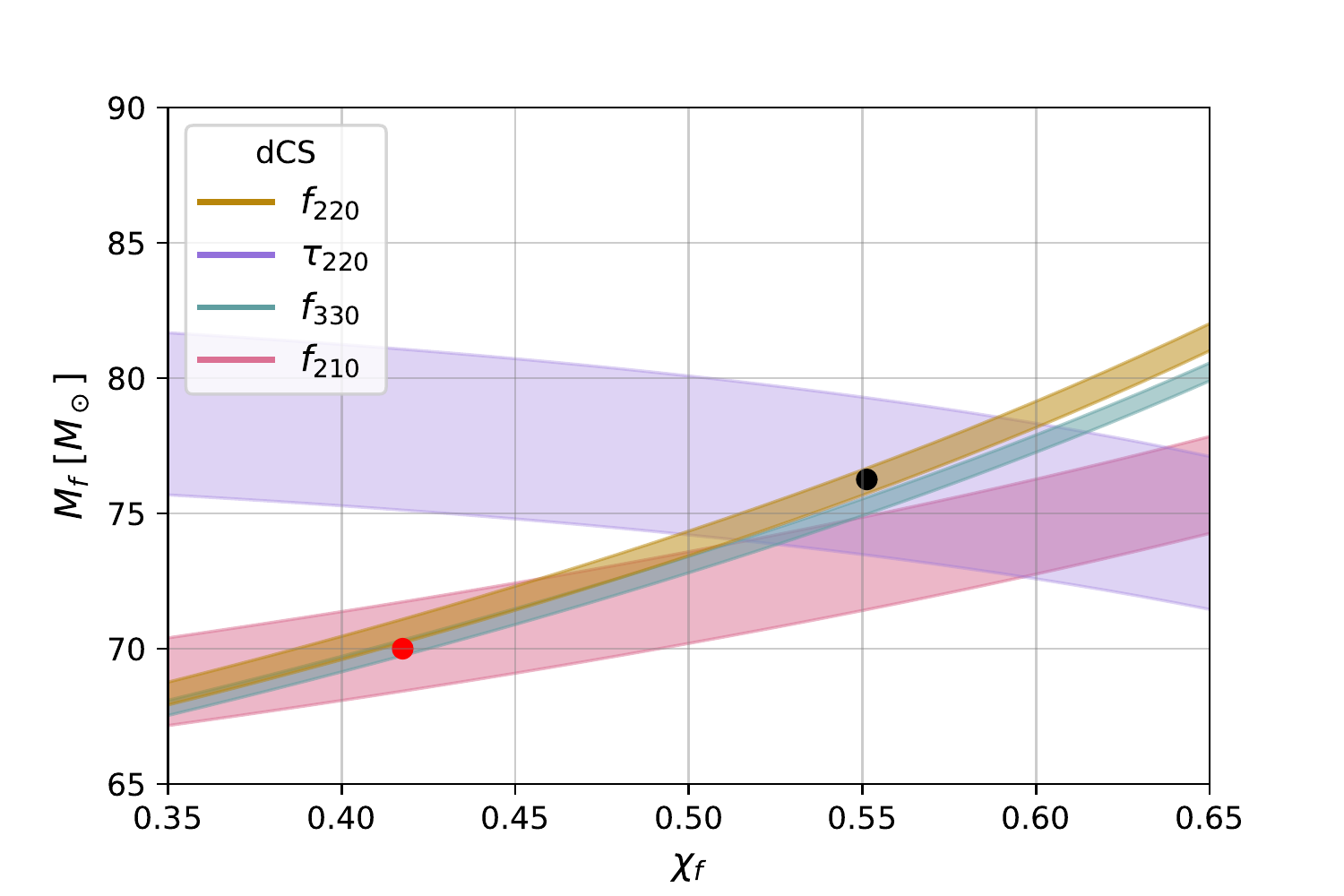}
    \caption{Projection of BH spectroscopy on the mass-spin plane for the Kerr-Newman (top) and dCS (bottom) spectra. We set $\rho_{\rm RD}=300$ and $q=5$. The bands correspond to the 90\% credible intervals. The black and red markers indicate the measurable and true values for mass and spin, respectively.}
    \label{fig:mass-spin-plot}
\end{figure}

To appreciate the quantitative results, we begin with illustrating the performance of BH spectroscopy for a) the best-case scenario, the dCS spectrum, and b) the worst-case scenario, the Kerr-Newman spectrum in Fig.~\ref{fig:mass-spin-plot}. For both cases, we use ringdowns with $\rho_{\rm RD} = 300$ and $q=5$. Specifically, we show the projections of the 90\% credibility bands of $f_{220}$, $\tau_{220}$, $f_{330}$ and $f_{210}$ on the mass-spin plane. The common region of intersection of $f_{220}$ and $\tau_{220}$ corresponds to the measurable mass and spin $\{\tilde{M}_f,\tilde{\chi}_f\}$ estimate (indicated by a black marker), while the true value of $\{{M}_f,{\chi}_f\}$ (indicated by a red marker) lies outside the intersection region. For the Kerr-Newman spectrum, all bands have a common intersection region, and therefore, BH spectroscopy fails to detect deviations from the GR BH Kerr QNMs. On the contrary, for the dCS spectrum, there is no common intersection region and BH spectroscopy can be used to identify that this spectrum is incompatible with GR at the 90\% confidence level.

Next, looking at the values of $\{\tilde{M}_f,\tilde{\chi}_f\}$ and of $\tilde{\delta} f_{lmn}$ in Tab.~\ref{tab:tilde}, we observe that $\{\tilde{M}_f,\tilde{\chi}_f\}$ does not differs significantly from $\{M_f,\chi_f\}$ and $\tilde{\delta} f_{lmn}\ll1$. Consequently, the results reported in Tab.~\ref{tab:fisher} for the GR Kerr BH spectra can be expected to approximately hold for the modified QNM spectra too. This allows to derive a back-of-the-envelope $\rho_{\rm RD}$ estimate required to distinguish various modified spectra from the GR Kerr BH spectrum. 
Further, we can use Eq.s~\eqref{eq:quantile:1}-\eqref{eq:quantile:2} to find the  $\rho_{\rm RD}$ necessary to exclude 0 from $P(\tilde{\delta}f_{330})$ at $90\%$ confidence.For instance, in the case of an EdGB spectrum, from \eqref{eq:quantile:1}  we observe that $Q^{lmn}_{\rm GR}=0.9$ corresponds to $p\approx1.64$. Inverting \eqref{eq:quantile:2} for $\rho_{\rm RD}=\kappa/\sigma$ and using the results in Tab.s~\ref{tab:tilde} and \ref{tab:fisher}, we get an approximate minimum value of $\rho_{\rm RD}$ as -- 
\begin{equation}
    \label{eq:rho_approx:1}
    \rho^{0.9}_{\rm RD}\approx
    \begin{cases}
    502 & (q=1.4)\\
    256 & (q=3)\\
    278 & (q=5)
    \end{cases}
\end{equation}

If we repeat this exercise for the dCS spectrum, we find
\begin{equation}
    \label{eq:rho_approx:2}
    \rho^{0.9}_{\rm RD}\approx
    \begin{cases}
    264 & (q=1.4)\\
    164 & (q=3)\\
    174 & (q=5)
    \end{cases}
\end{equation}
Note that it is only for the sake of demonstrating a fast and easy approximate estimation that we derive \eqref{eq:rho_approx:1} using the values in Tab.~\ref{tab:fisher}, which assumed a GR Kerr BH spectrum. However, there is no fundamental obstruction to being fully consistent by applying Eq.s~\eqref{eq:quantile:1} and \eqref{eq:quantile:3} using the covariance matrices from the modified spectra.

\begin{figure*}[p]
    \centering
    \includegraphics[width=0.43\textwidth]{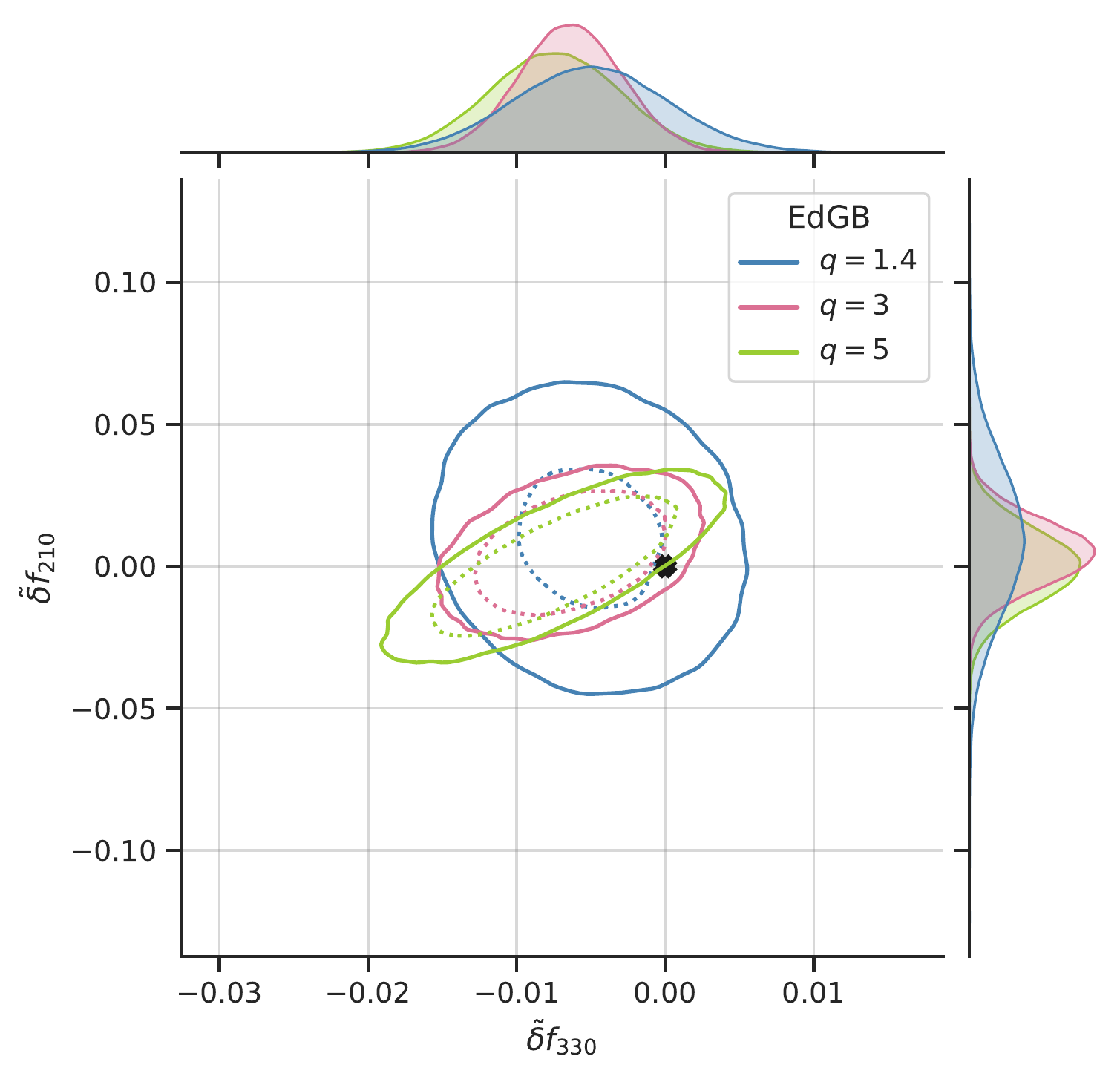}
    \includegraphics[width=0.43\textwidth]{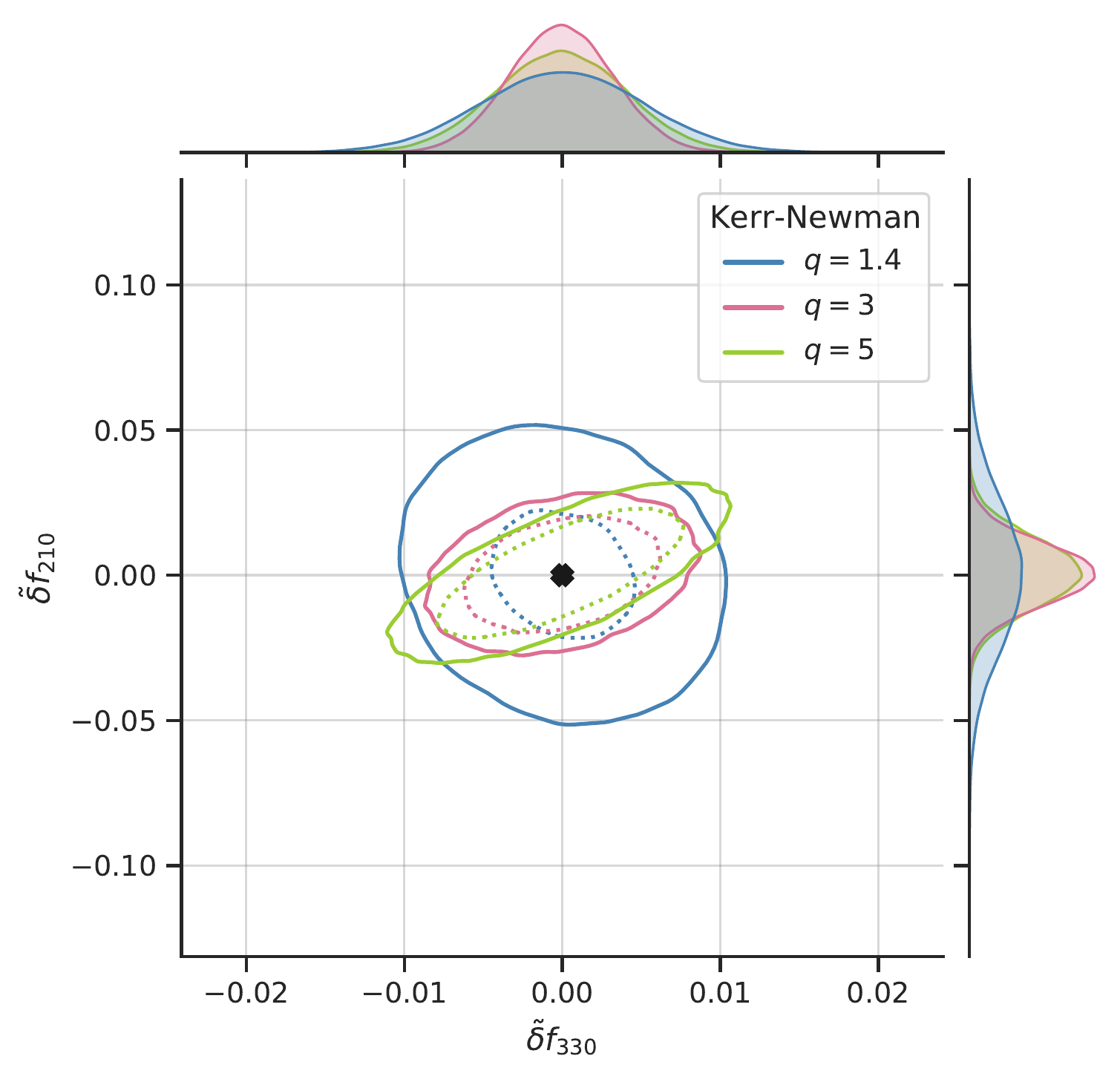}\\
    \includegraphics[width=0.43\textwidth]{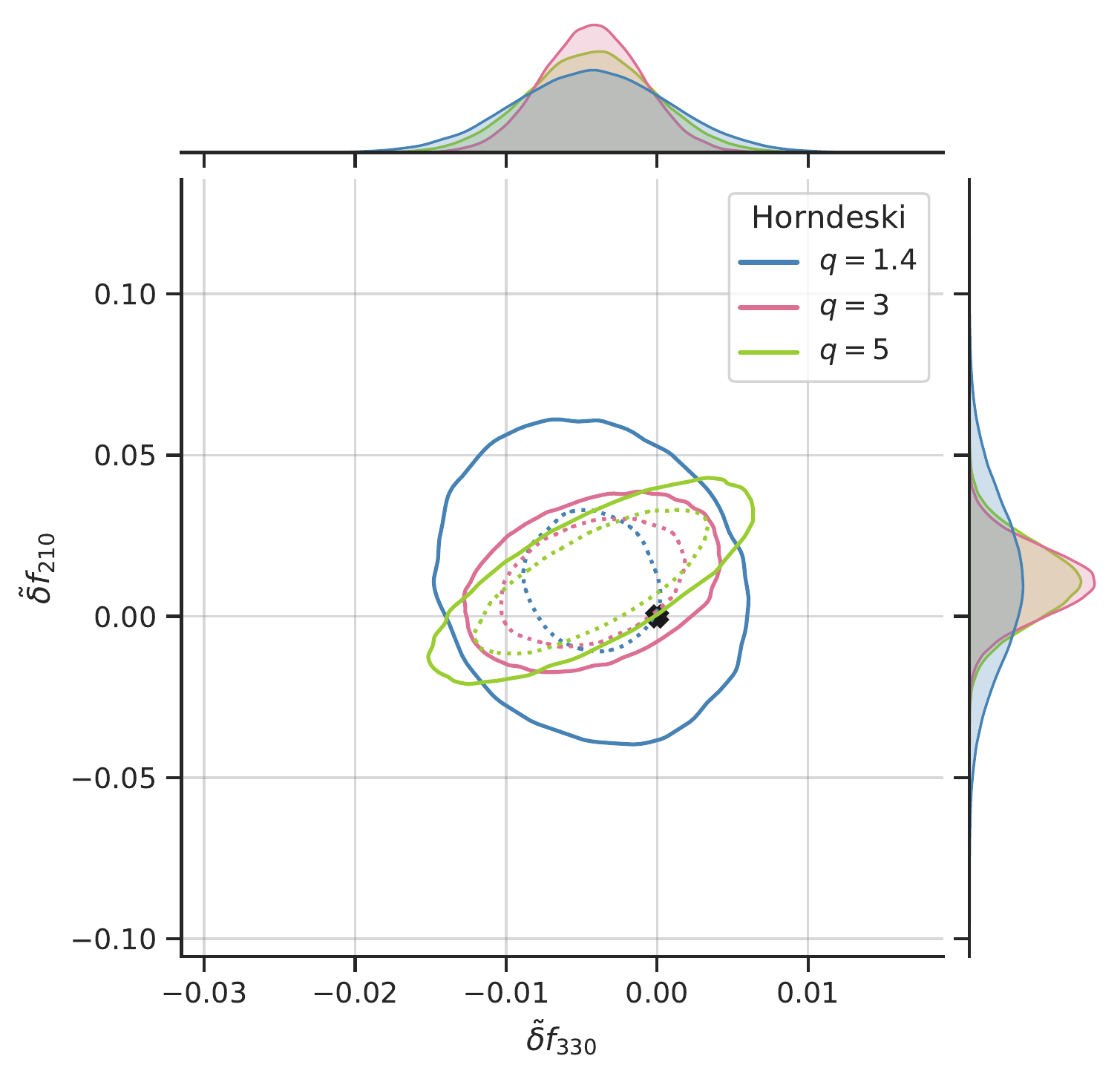}
    \includegraphics[width=0.43\textwidth]{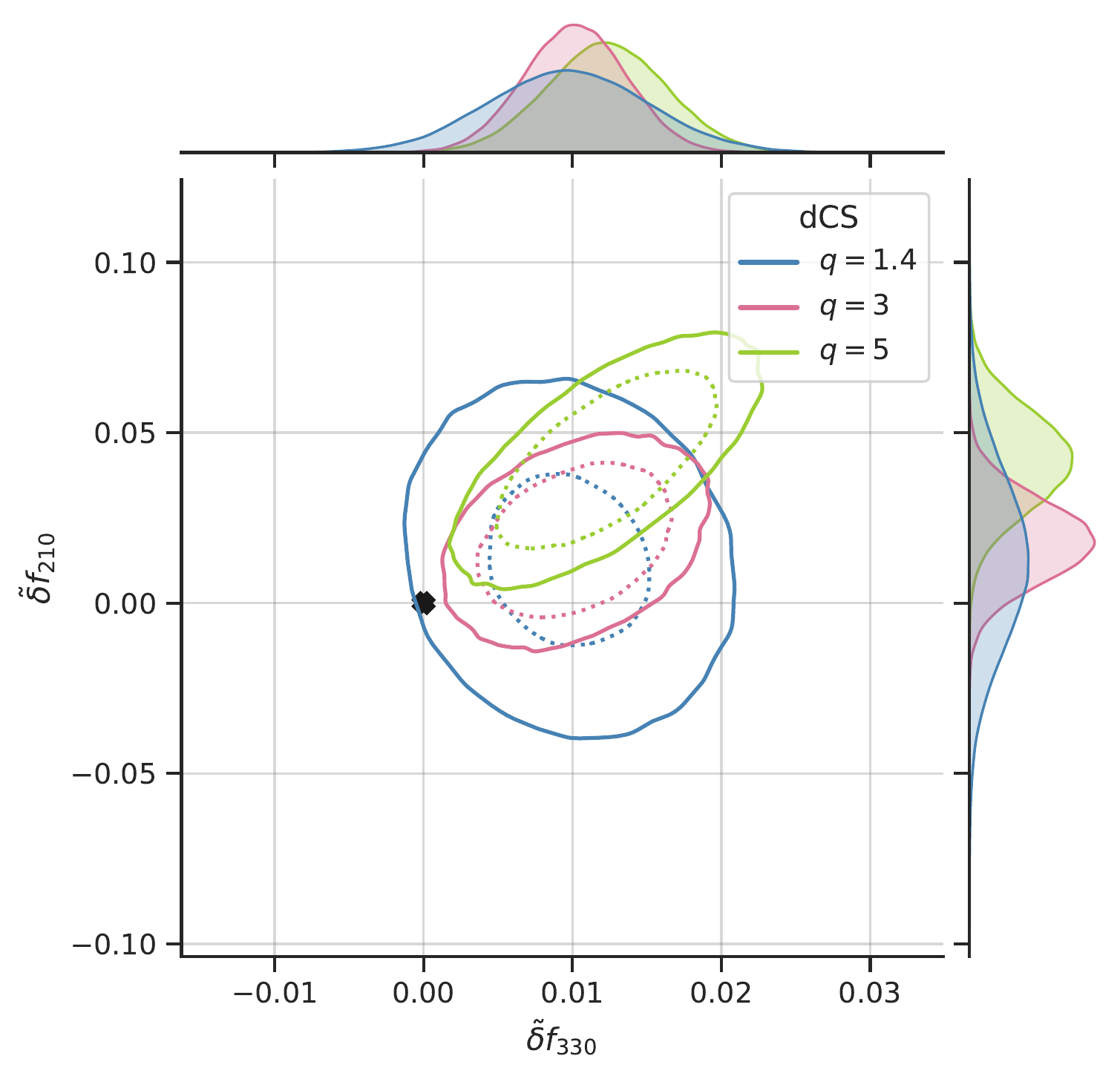}\\  
    \includegraphics[width=0.43\textwidth]{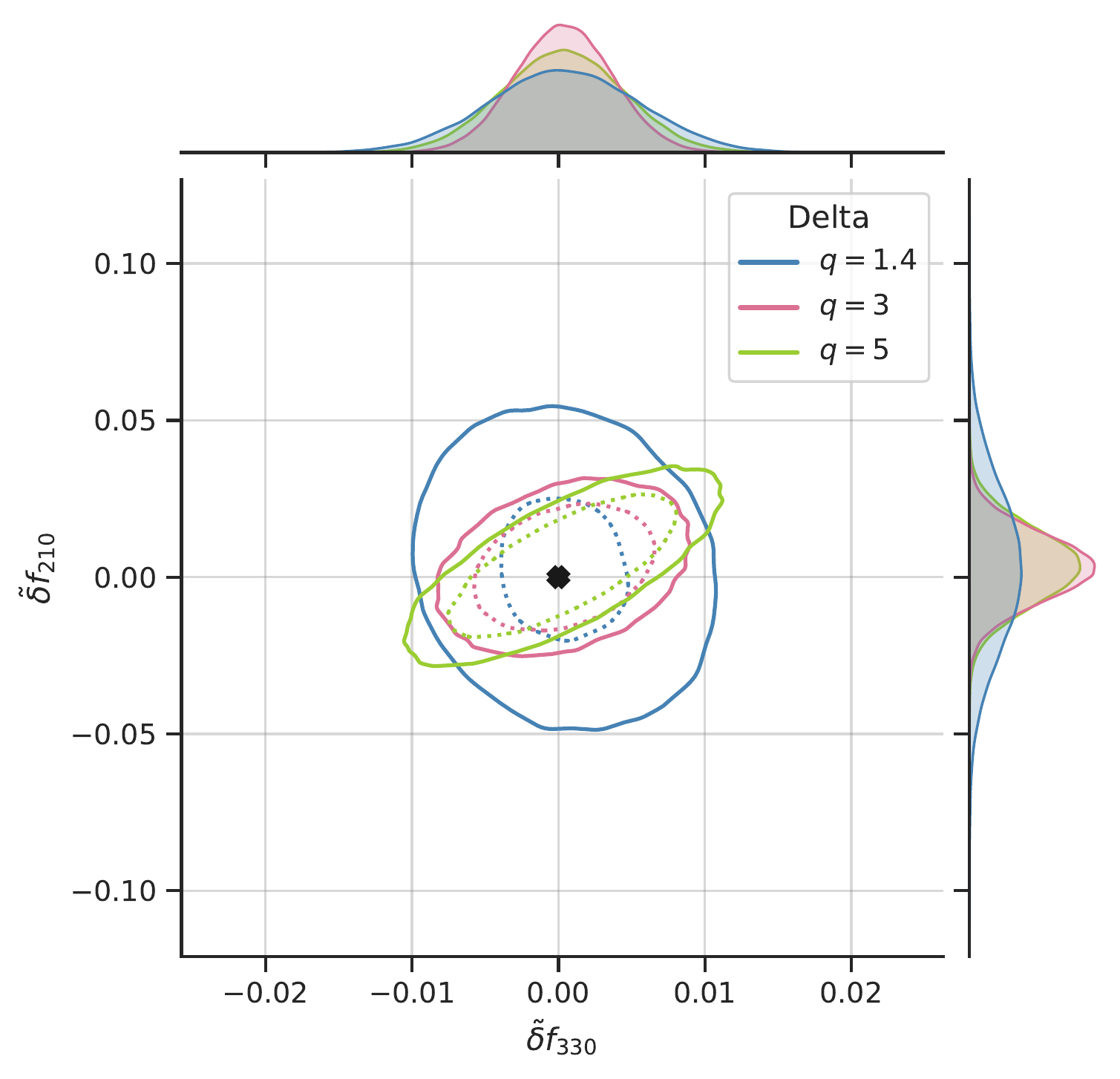}
    \includegraphics[width=0.43\textwidth]{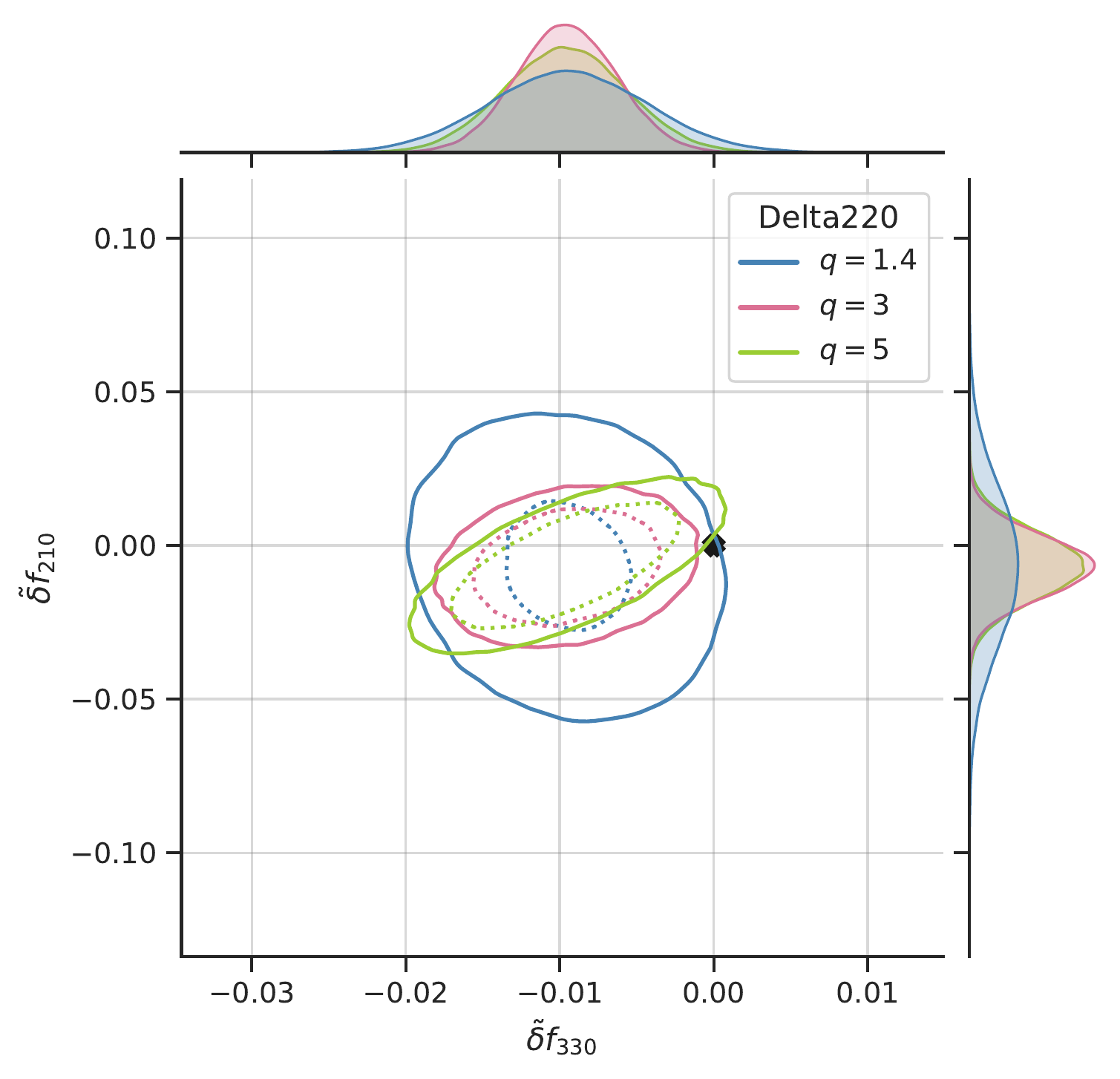}
    \caption{Density plots of $\{\tilde{\delta}f_{330},\tilde{\delta}f_{210}\}$ for the modified spectra considered in Sec.~\ref{sec:spectra} and different mass ratios. Solid and dashed lines indicate the $90\%$ and $50\%$ probability contours respectively. We assume a true injected mass $M_f=70 M_\odot$ and vanishing initial spins (see Tab.~\ref{tab:tilde} for the values of the corresponding effective measurable masses and spins). We set to $\rho_{\rm RD}=300$ to clearly display the different trends of the spectra in distinguishing deviations from GR. The null hypothesis under test is denoted by a black marker.}
    \label{fig:corner}
\end{figure*}

\begin{figure*}[p]
    \centering
    \includegraphics[width=0.32\textwidth]{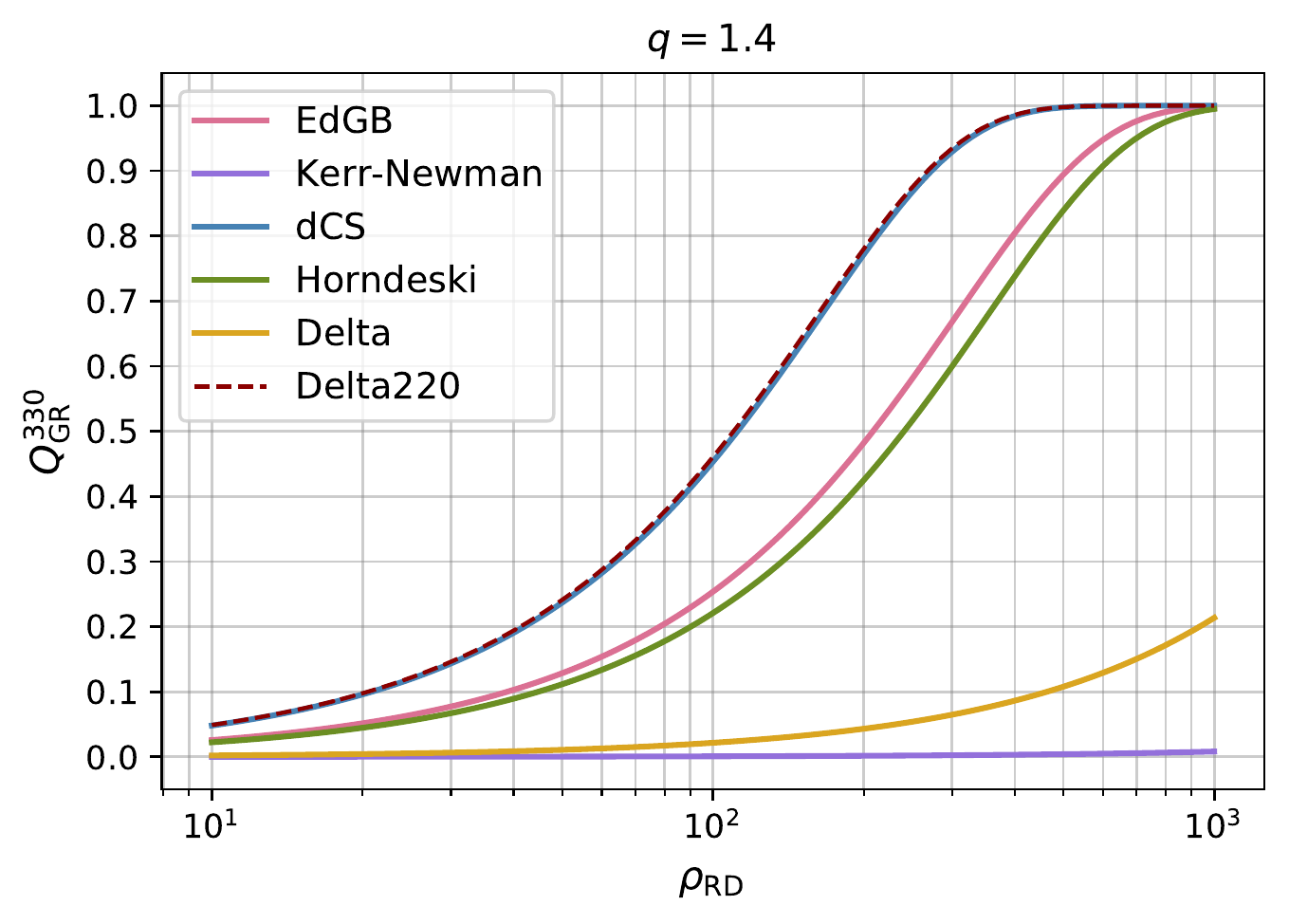}
    \includegraphics[width=0.32\textwidth]{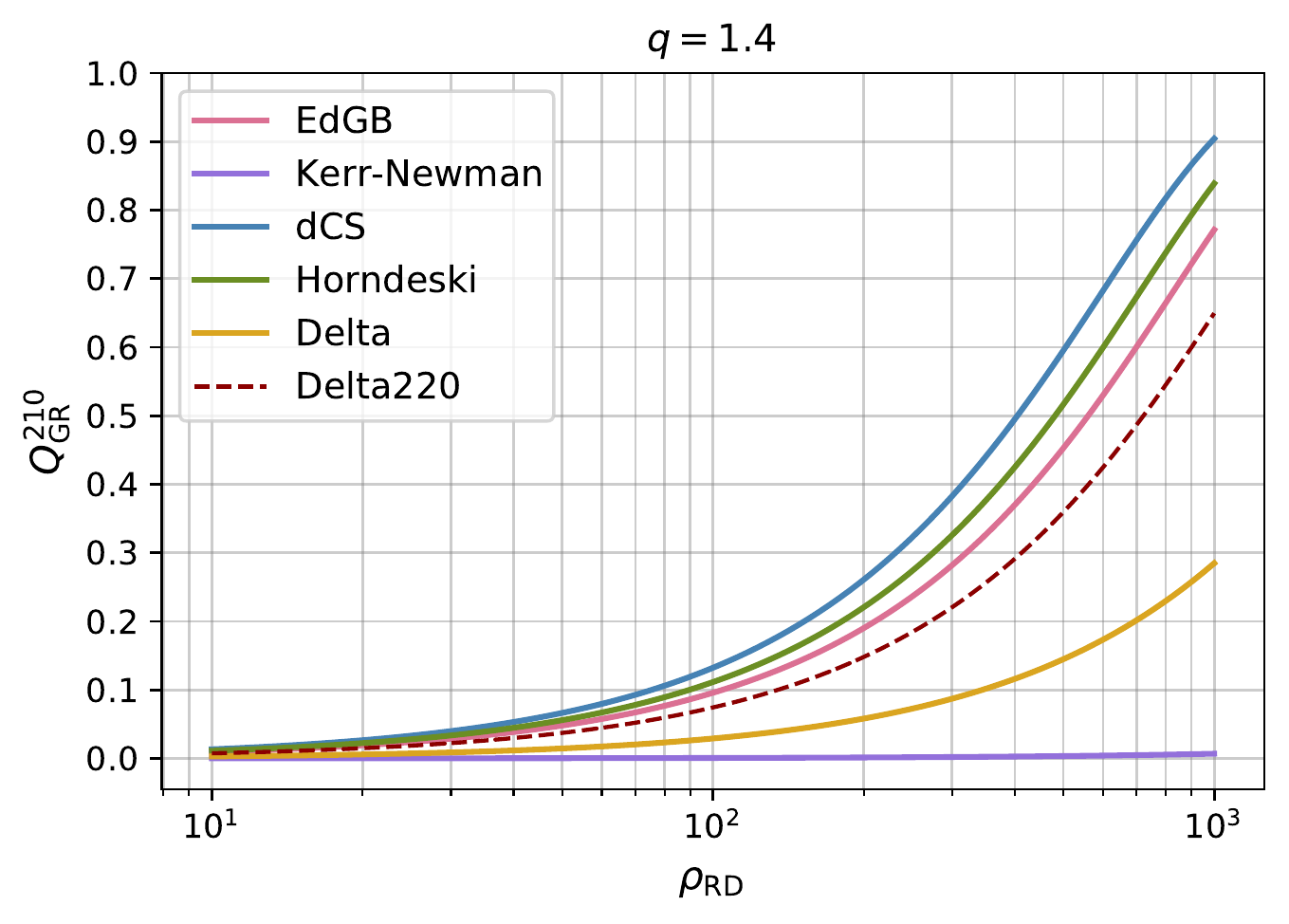}
    \includegraphics[width=0.32\textwidth]{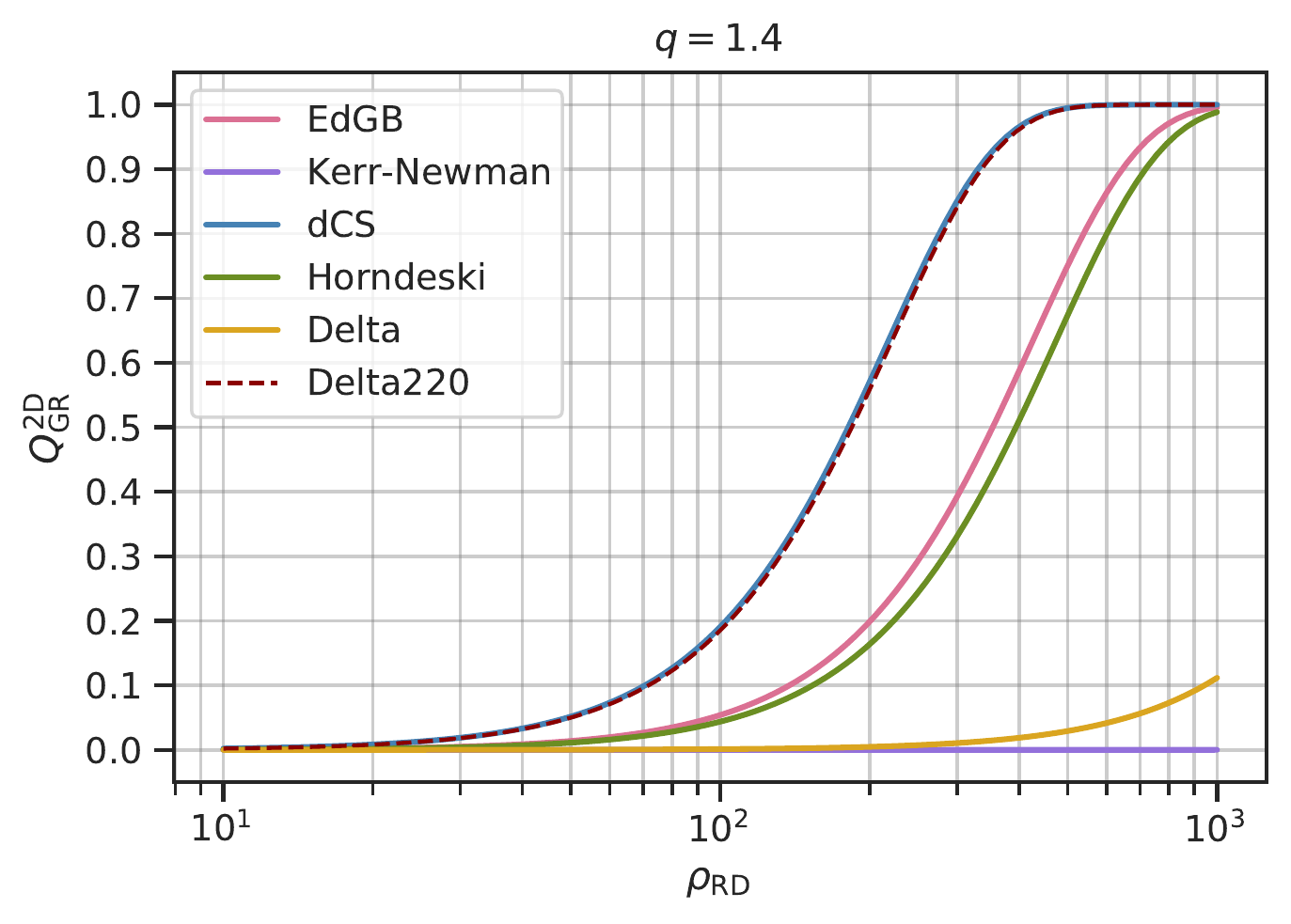}\\
    \includegraphics[width=0.32\textwidth]{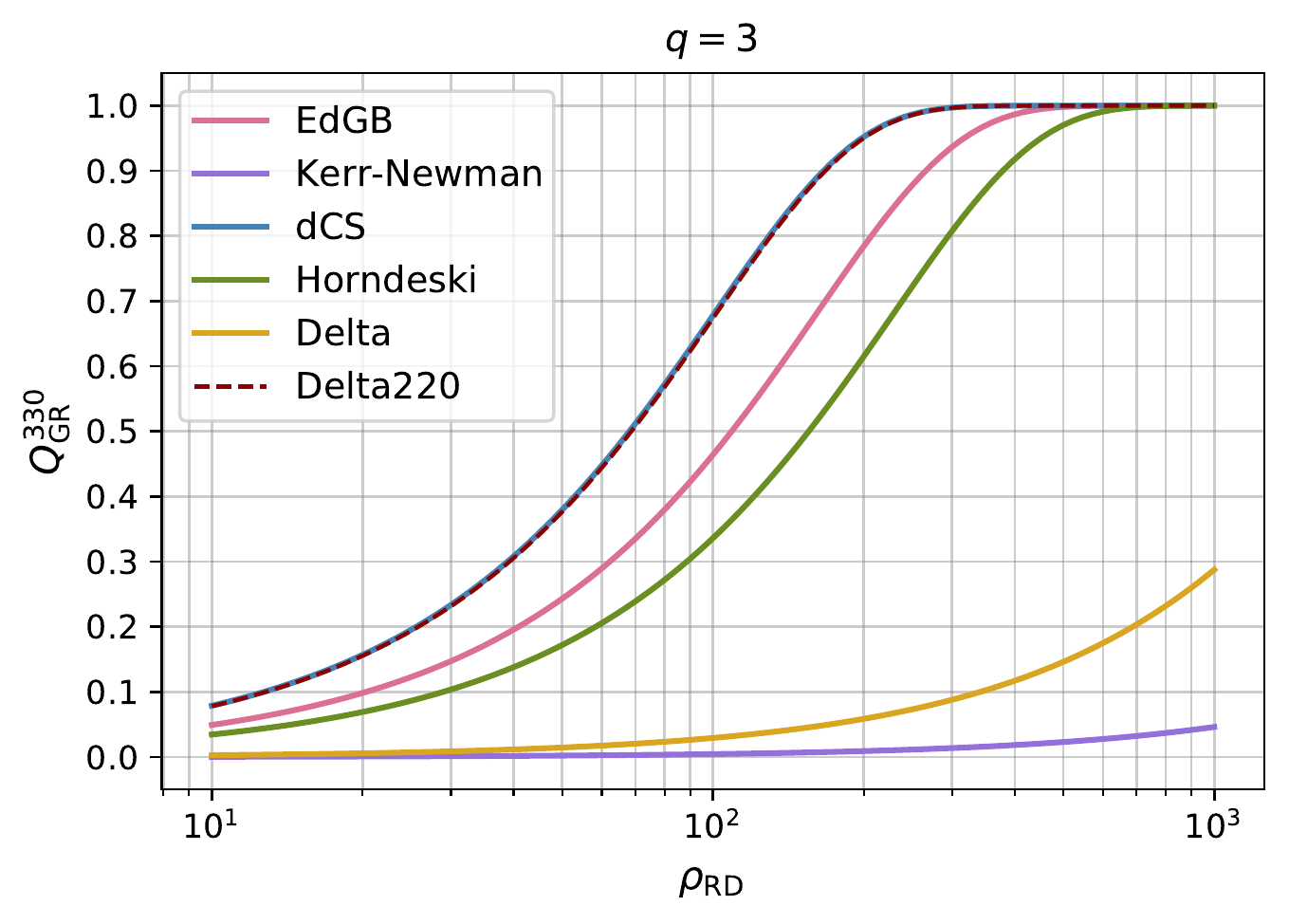}
    \includegraphics[width=0.32\textwidth]{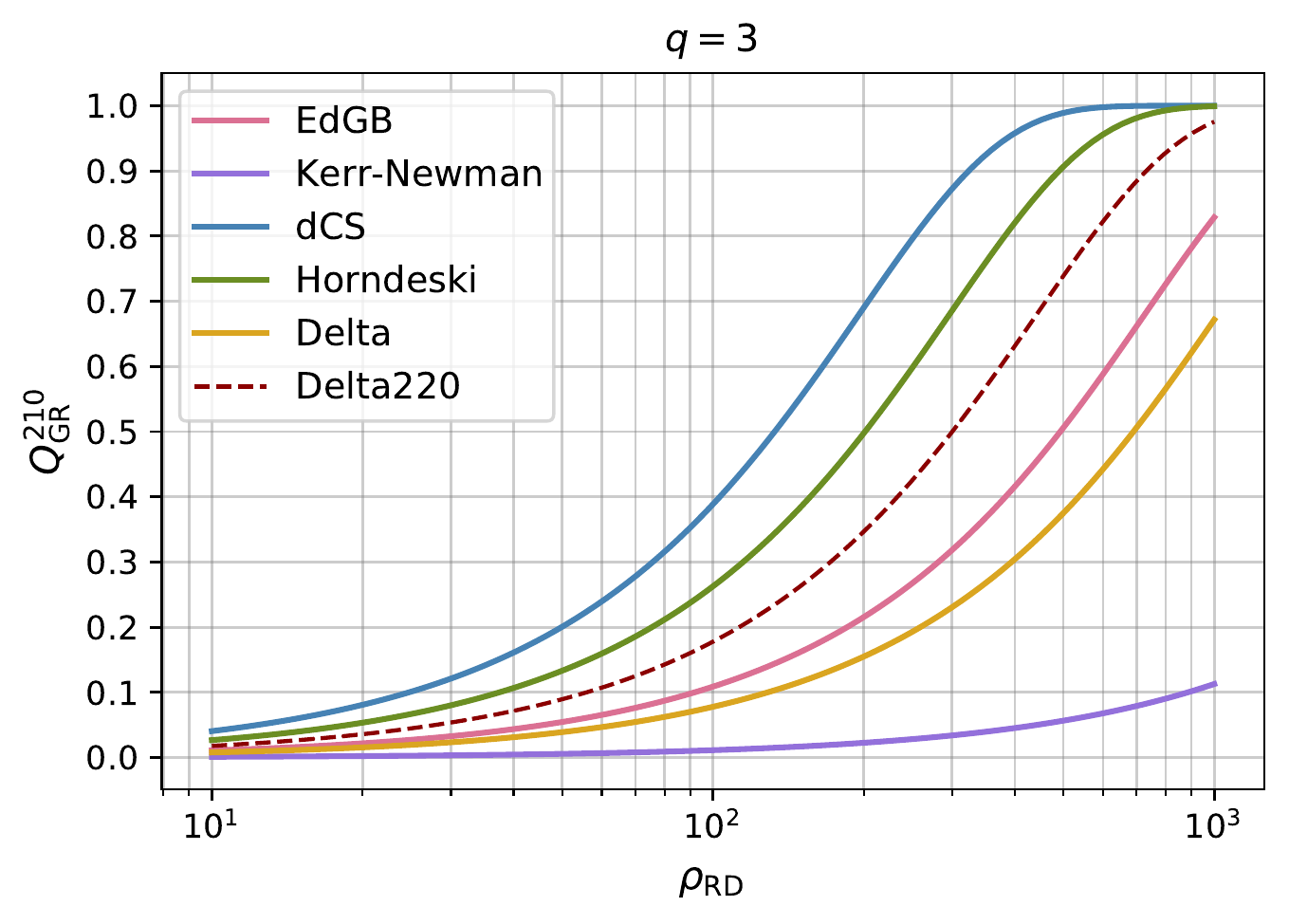}
    \includegraphics[width=0.32\textwidth]{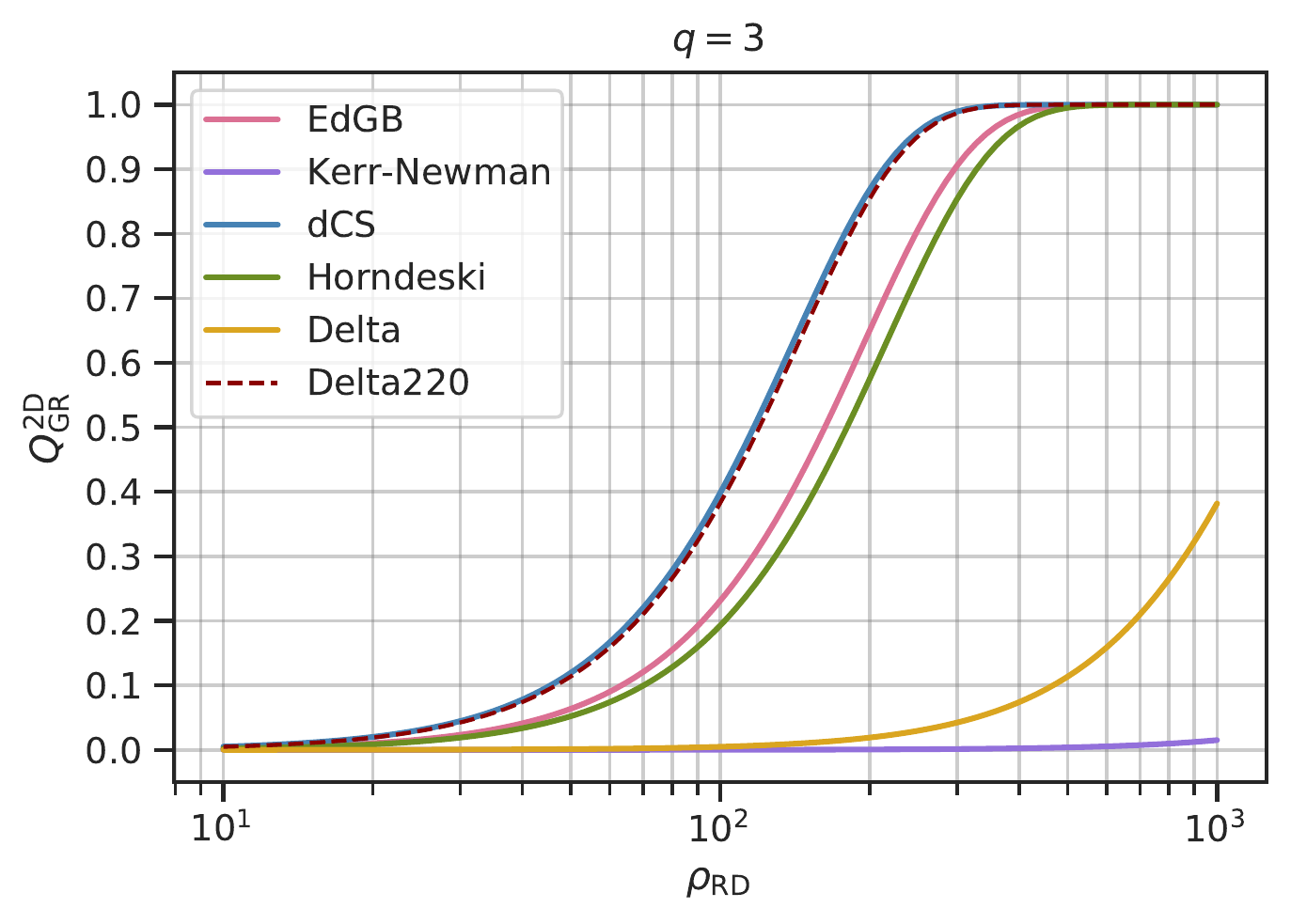}\\
    \includegraphics[width=0.32\textwidth]{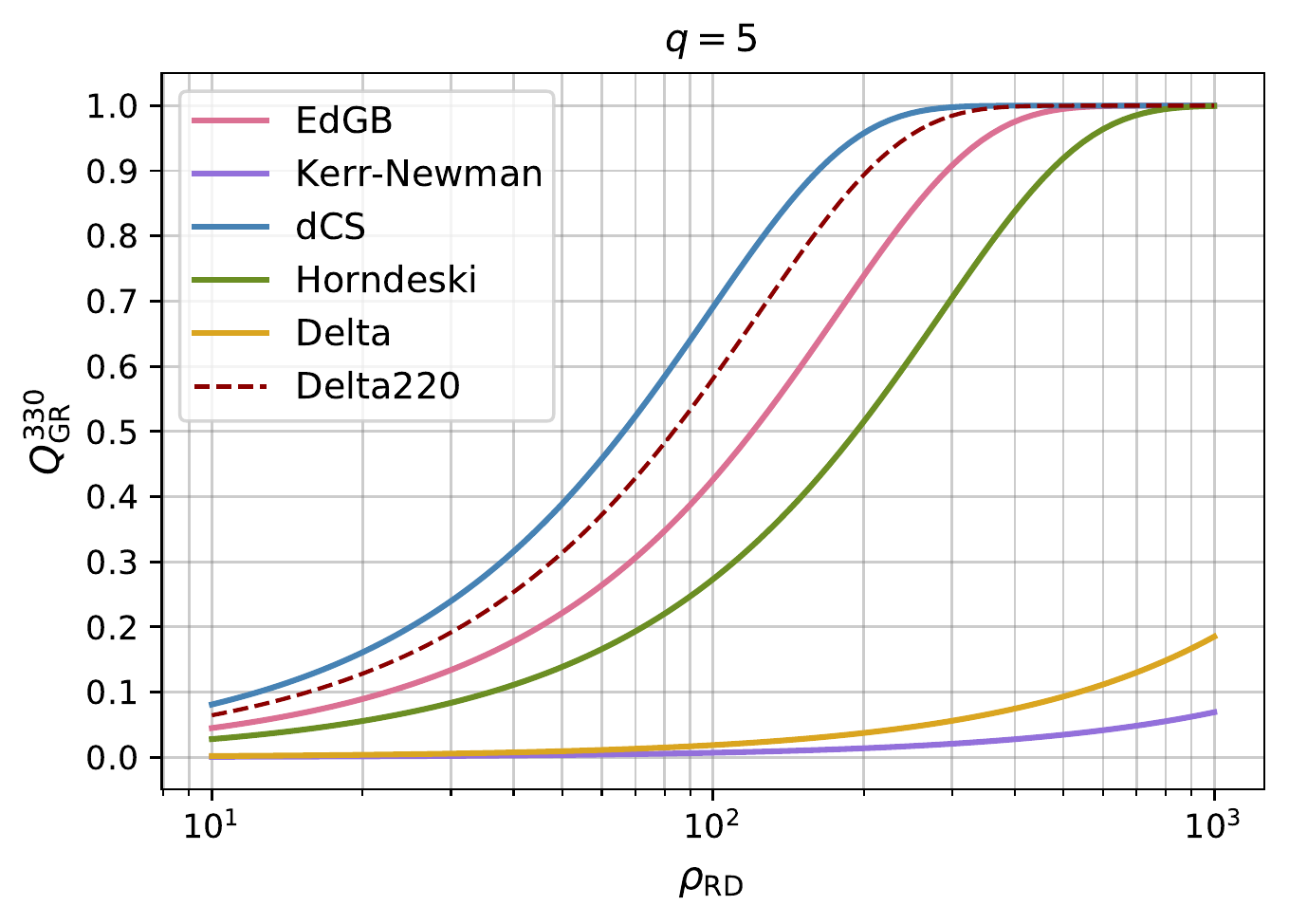}
    \includegraphics[width=0.32\textwidth]{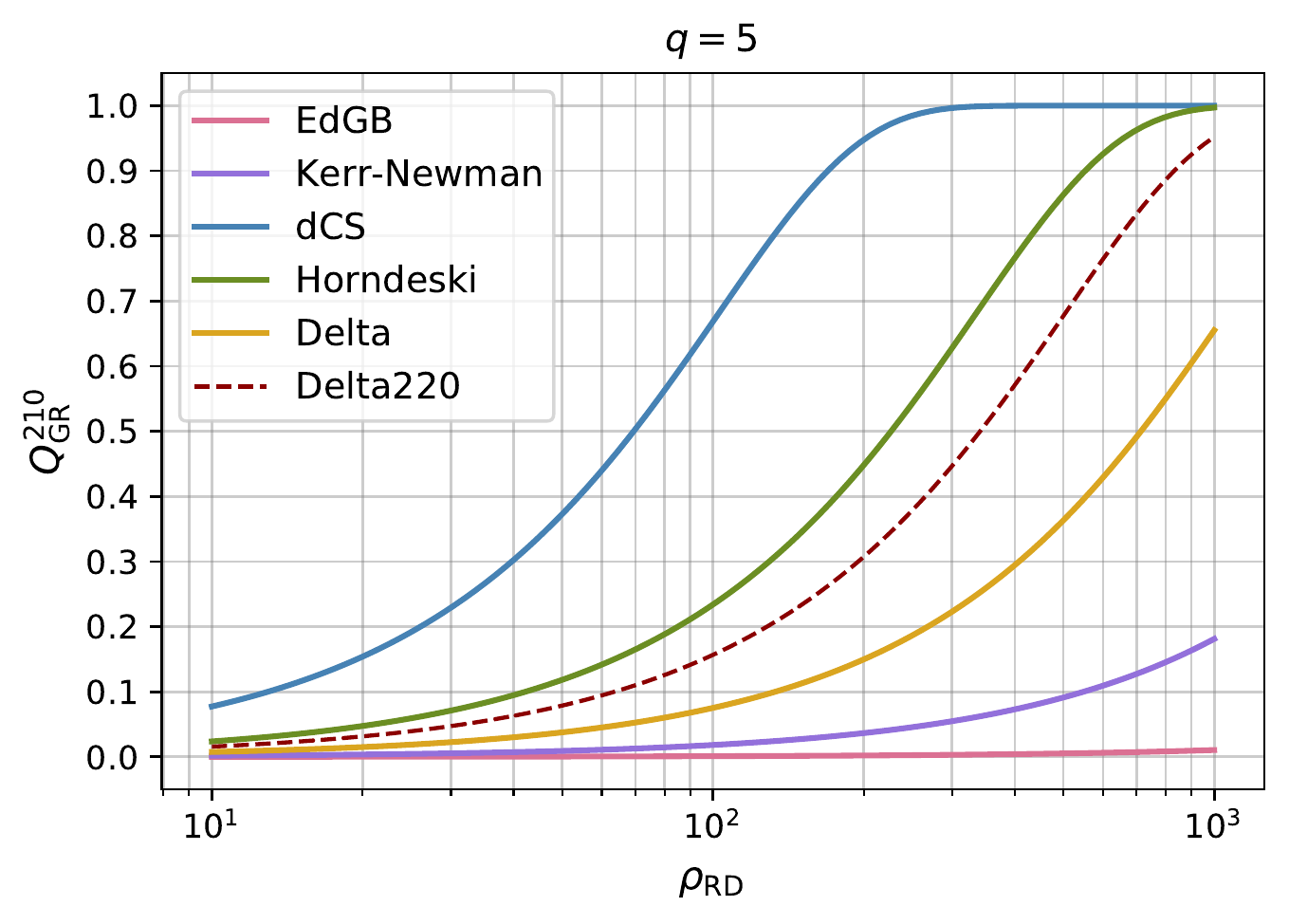}
    \includegraphics[width=0.32\textwidth]{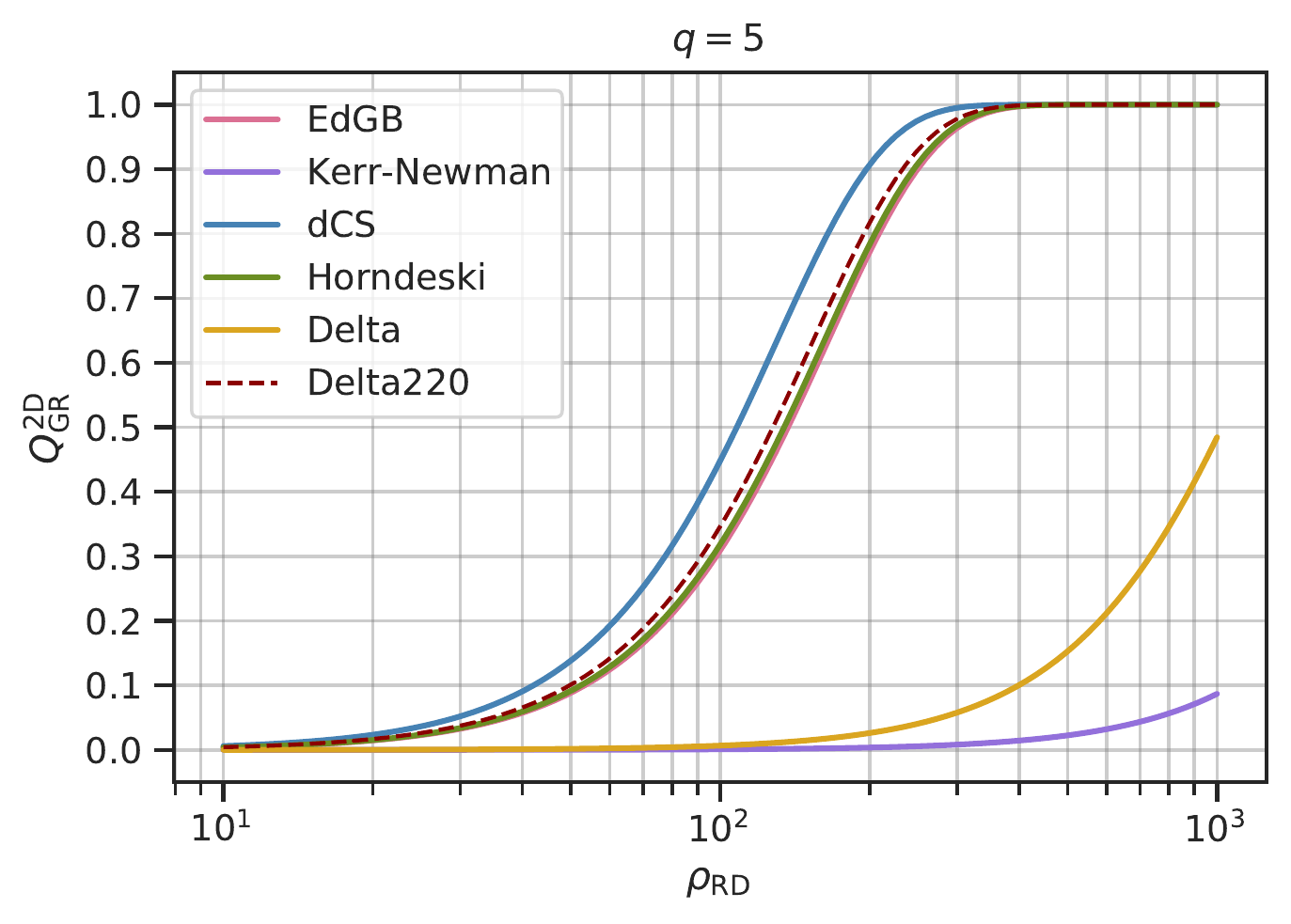}\\
    \caption{Values of the GR quantile $Q_{\rm GR}$ as a function of the $\rho_{\rm RD}$. We consider quantiles for the marginalised single parameter posteriors $P(\tilde{\delta} f_{330})$ (leftmost column) and $P(\tilde{\delta} f_{210})$ (central column) and for the 2D joint posterior $P(\tilde{\delta} f_{330},\tilde{\delta} f_{210})$ (rightmost column). The quantiles are computed using the exact expressions \eqref{eq:quantile:1} and \eqref{eq:quantile:3}.}
    \label{fig:qgr}
\end{figure*}

In Fig.~\ref{fig:corner}, we depict marginalized 2-d posterior estimates of $\tilde{\delta} f_{330}$ and $\tilde{\delta} f_{210}$ for all the modified ringdowns studied. Here again, we use ringdowns with $\rho_{\rm RD} = 300$ to illustrate $Q_{\rm GR}$ as a measure to distinguish a modified QNM spectrum from the GR Kerr BH spectrum. When the posterior distributions are compatible with $(0,0)$ (indicated with a black dot), the QNM spectra for the $f_{330}$ and $f_{210}$ are compatible with the GR Kerr BH spectrum. Note also that the shape of the contours changes with $q$ (and that the contours do not trivially shrink with changing in $q$). This foreshadows the non-trivial dependence of $Q_{\rm GR}$ on $q$ which will be further emphasised in fig \ref{fig:qgr}. We can identify the modifications to various level of confidence --  $Q_{\rm GR}^{\rm 2D} = \{0.33, 0.85, 0.97 \}$ for Horndeski, $Q_{\rm GR}^{\rm 2D} = \{0.39, 0.91, 0.96 \}$  for EdGB, $Q_{\rm GR}^{\rm 2D} = \{0.85, 0.99, 1 \}$ for dCS for $q= \{1.4, 3,5 \}$ respectively.  We can contrast this to the case of Kerr-Newman where we do not expect to detect any deviations  -- we have $Q_{\rm GR}^{\rm 2D} = \mathcal{O}(10^{-6})$ for $q=1.4$,  $\mathcal{O}(10^{-3})$ for $q=3$ and $\mathcal{O}(10^{-2})$ for $q=5$. We will see in Fig.~\ref{fig:qgr} that this is true even at very high $\rho_{\rm RD}$.

Finally, in Fig.~\ref{fig:qgr}, we present our main results on $\rho_{\rm RD}$ required to identify deviations from the GR Kerr QNM spectrum for the various modification schemes. The first two columns show $Q_{\rm GR}$ computed using a 1-d posterior distribution of $f_{330}$ and   $f_{210}$, respectively, and the rightmost column corresponds to the 2-d joint posterior distribution on $f_{330}$ and $f_{210}$. Again, we study systems corresponding to three mass ratios: in the first row, we study the near-equal-mass scenario of $q = 1.4$; in the second row, we study $q=3$ and in the last row, we study $q=5$. 

Let us examine the $\rho_{\rm RD}$ at which $Q_{\rm GR}\geq 0.9$ for all the modified QNMs considered here. We observe that  $\rho_{\rm RD}$ required to distinguish the modified QNMs from GR Kerr BH spectrum depends acutely on the details of the modifications. Modification of the spectra, such as that predicted by Kerr-Newman and Delta, cannot be confidently differentiated from the GR spectrum, even when $\rho_{\rm RD}=10^{3}$. This happens because these modifications can be approximated by the GR by suitably selecting the values of $\tilde M_{f}$ and $\tilde \chi_{f}$. In other words, these modified spectra are highly degenerate with the GR Kerr BH QNM in the mass-spin space. This can be quantitatively seen in Fig.~\ref{fig:spectra}. For instance, $\tilde{\delta} f_{lmn}$ for these modified spectra is an order of magnitude smaller than $\tilde{\delta} f_{lmn}$ for other modifications such as dCS or EdGB. In contrast to this, modifications predicted by the EdGB and Horndeski theories can be distinguished confidently from the GR at a $\rho_{\rm RD} \in [250, 10^{3}]$. Furthermore, the dCS or Delta220-like QNM spectra could be confidently identified with an even lower $\rho_{\rm RD} \in [250, 400]$. 

We highlight two non-trivial trends from Fig.~\ref{fig:qgr} that we observe.

\begin{itemize}
\item While all of them are valid measures to identify modifications in a QNM spectrum, we find that the ability to distinguish a modified theory from GR using 1-d $Q_{\rm GR}^{330}$, $Q_{\rm GR}^{210}$ and 2-d joint  $Q_{\rm GR}^{\rm 2D}$ are fairly different and depends on theory. While modifications in the spectra predicted by dCS and Delta220 benefit by using $Q_{\rm GR}^{330}$, Kerr-Newmann and Delta are more distinguishable using $Q_{\rm GR}^{210}$. Furthermore, certain types of modifications will only manifest in a detectable fashion in one of the $Q_{\rm GR}$ measures. For example, EdGB spectra with $\rho_{\rm RD} \leq 10^{3}$ cannot be confidently distinguished if we used $Q_{\rm GR}^{210}$ while it can be using $Q_{\rm GR}^{330}$ or $Q_{\rm GR}^{\rm 2D}$.

\item The \emph{distinguishability} of a given modified spectrum does not have a  monotonic behaviour with $q$. This occurs because $q$ dictates both the mode excitation amplitudes and the final spin of the remnant BH, and the interplay between the two  produces the non-monotonic trend we see here. For instance, at a given $\rho_{\rm RD}$, $Q_{\rm GR}^{330}$ for dCS increases monotonically with $q$, while $Q_{\rm GR}^{330}$ for EdGB and Horndeski spectra seem to perform better for the case of $q=3$ than for $q=1.4$ or $5$. It is also worth noting that the performance of $Q_{\rm GR}^{330}, Q_{\rm GR}^{210}$ and $Q_{\rm GR}^{\rm 2D}$ for different theory can differ with $q$. 
\end{itemize}
\section{Discussion and conclusion}
\label{sec:dis}
We performed a comprehensive study on the ability of BH spectroscopy to distinguish a modified spectra from a GR Kerr BH QNMs. We studied theory-motivated modifications for QNM spectra that were publicly available -- specifically, EdGB, Kerr-Newman, Horndeski and dCS theory, and for two phenomenologically modified QNM spectra -- Delta and Delta220. To investigate $\rho_{\rm RD}$ necessary to distinguish these modified spectra from a GR Kerr BH spectrum, we assessed the performance of BH spectroscopy using a Fisher information matrix formalism. The ringdowns with the modified spectra are generated such that in each case $f_{220}$ deviates from the corresponding GR Kerr QNM by $1 \% $. 

First, we re-iterate that we can only measure the effective deviation parameters $\tilde{\delta} f_{lmn}$ and not the absolute deviation $\delta f_{lmn}$ as the mass and the spin estimation can be chosen suitably to compensate for the deviations of the QNMs. Further, we show that in many theories $\tilde{\delta} f_{lmn}$ significantly differs from $\delta f_{lmn}$. Therefore, to study the ability BH spectroscopy to distinguish modified spectra, the framework must be setup using the measurable effective deviation parameters $\tilde{\delta} f_{lmn}$ .

We found that $\rho_{\rm RD}$ necessary to distinguish a modified spectrum from the GR Kerr one depends on the details of the modification and on the mass ratio. Further, we used three different measures that quantify the amount by which: a) the 1-d posterior estimate of $\tilde{\delta} f_{330}$ excludes 0 a.k.a.~$Q_{\rm GR}^{330}$, b) the 1-d posterior estimate of $\tilde{\delta} f_{210}$ excludes 0 a.k.a.~$Q_{\rm GR}^{210}$ and c) the 2-d joint posterior estimation of $\tilde{\delta} f_{330}-\tilde{\delta} f_{210}$ excludes (0,0) a.k.a.~$Q_{\rm GR}^{\rm 2D}$. Depending on the spectrum and the magnitudes of $\tilde{\delta} f_{lmn}$, the performance of $Q_{\rm GR}^{330}, Q_{\rm GR}^{210}$ and $Q_{\rm GR}^{\rm 2D}$ can vary significantly. Roughly, we find that  a $\rho_{\rm RD} \geq 150$ is required to identify deviations from GR at a 90\% credibility level. This range of $\rho_{\rm RD}$ is attainable with the next generation detectors and BH spectroscopy will be a powerful tool to constrain GR as well as for identifying modifications to Kerr BH spectrum. 

In previous works \cite{Bhagwat:2021kwv,Bhagwat:inprep}, we studied the measurability of QNM parameters, that is, the statistical uncertainty with which a QNM mode parameter can be estimated from a signal to assess the landscape of BH spectroscopy with a next-generation detector. However, we note here that the subdominant QNM mode with the best measurability may not always be the optimal mode for distinguishing a modified QNM spectrum from a GR Kerr BH QNM spectrum. The best mode to identify a departure from GR Kerr spectrum depends on the interplay between measurablity and the details of modification of the spectra. For instance, measurablity of $f_{330}$ is better than $f_{210}$ in a non-spinning binary BH ringdown. However, if the compact object is a Kerr-Newman BH instead of a Kerr BH, $f_{330}$ measured in the ringdown would be compatible with the Kerr BH even for ringdown with $\rho_{\rm RD} \sim 10^{3}$. The modifications in the spectra in this case can be observed predominantly by looking at $f_{210}$. Therefore, using a setup that uses information in all measurable QNM mode parameters is optimal while looking for departure from the GR Kerr spectrum in a ringdown signal. Greater the number of QNM mode parameters measured, greater the chances that we can distinguish a modified QNM spectra from the GR Kerr spectra.
 
Furthermore, among the modified QNM spectra we have studied, we see the theories separate out into two groups :
\begin{itemize}
\item EdGB, Horndeski, dCS, Delta220: For these spectra, a GR Kerr ringdown can be excluded at $90\%$ confidence level with a SNR $\rho^{0.9}_{\rm RD}\in[150,500]$. The deviations manifest more prominently in the $(3,3,0)$ mode and the $(2,1,0)$ mode offers a relatively poor constraints.
\item Kerr-Newman, Delta: here the deviations manifest almost exclusively in the $(2,1,0)$ mode, with $Q_{\rm GR}^{210}$ returning the most performative no-hair test. However, the $\rho_{\rm RD}$ required to exclude 0 at 90\% confidence is so high that these deviations are indistinguishable, even at $\rho_{\rm RD}=1000$.
\end{itemize}

Finally, works such as this are essential for gauging the potential of ringdown-based tests of gravity; but unfortunately, we are restricted by publicly available modified QMN spectra. While it is difficult to solve the QNM spectra in modified theories of gravity, to adequately prepare for test of GR with next-generation detectors including designing the implementation of BH spectroscopy, we feel that the field would benefit substantially from investing effort in this direction.
\begin{acknowledgments}
We thank L. Pierini for clarifications about the work in \cite{Pierini:2022eim}. We thank E. Berti, L. Gualtieri, A. Maselli and P. Pani for fruitful discussions.
C.P.~is supported by European Union's H2020 ERC Starting Grant No. 945155--GWmining and by Cariplo Foundation Grant No. 2021-0555.
S.B. would like to acknowledge the UKRI Stephen Hawking Fellowship funded by the Engineering and Physical Sciences Research Council (EPSRC) with grant reference number EP/W005727 for support during this project. 
\end{acknowledgments}
\bibliography{refs.bib}
\end{document}